\documentclass[journal=jctcce,manuscript=article]{achemso}

\usepackage[version=3]{mhchem} 
\usepackage[T1]{fontenc}
\usepackage[utf8]{inputenc}
\usepackage{xcolor}
\usepackage{subcaption}
\usepackage{bm}
\usepackage{placeins} 
\usepackage{pdfpages}

\title{Bridging Atomistic and Mesoscale Lithium Transport via Machine-Learned Force Fields and Markov State Models}

\author{Muhammad Nawaz Qaisrani}
\affiliation{Ilmenau University of Technology, Theoretical Solid State Physics,\\ Weimarer Straße 32, 98693 Ilmenau, Germany}

\author{Christoph Kirsch}
\affiliation{Martin-Luther-University Halle-Wittenberg, Institute of Chemistry, Theoretical Chemistry,\\ Von-Danckelmann-Platz 4, 06120 Halle (Saale), Germany}

\author{Aaron Fl\"ototto}
\affiliation{Ilmenau University of Technology, Theoretical Solid State Physics,\\ Weimarer Straße 32, 98693 Ilmenau, Germany}

\author{Jonas H{\"a}nseroth}
\affiliation{Ilmenau University of Technology, Theoretical Solid State Physics,\\ Weimarer Straße 32, 98693 Ilmenau, Germany}

\author{Jules Oumard}
\affiliation{Ilmenau University of Technology, Theoretical Solid State Physics,\\ Weimarer Straße 32, 98693 Ilmenau, Germany}

\author{Daniel Sebastiani}
\affiliation{Martin-Luther-University Halle-Wittenberg, Institute of Chemistry, Theoretical Chemistry,\\ Von-Danckelmann-Platz 4, 06120 Halle (Saale), Germany}

\author{Christian Dre{\ss}ler}
\email{christian.dressler@tu-ilmenau.de}
\affiliation{Ilmenau University of Technology, Theoretical Solid State Physics,\\ Weimarer Straße 32, 98693 Ilmenau, Germany}

\abbreviations{MLFFs, LIBs, Li\textsubscript{x}Si\textsubscript{y}, MD, AIMD, MSMs}
\keywords{Machine-learned force field, Lithium-ion batteries, Lithium silicides, Molecular dynamics, Ab initio molecular dynamics, Markov state models}

\begin{document}

\begin{abstract}
Lithium diffusion in solid-state battery anodes occurs through thermally activated hops between metastable sites often separated by large energy barriers, making such events rare on \textit{ab initio} molecular dynamics (AIMD) timescales.
Here, we present a bottom-up multiscale workflow that integrates AIMD, machine-learned force fields (MLFFs), and Markov state models (MSMs) to establish a quantitatively consistent link between atomistic hopping mechanisms and mesoscale transport. MLFFs fine-tuned on AIMD reference data retain near-DFT accuracy while enabling large-scale molecular dynamics simulations extending to tens of nanoseconds. These extended trajectories remove the strong finite-size bias present in AIMD and yield diffusion coefficients in excellent agreement with experiment. Furthermore, from these long MLFF trajectories, we obtain statistically converged lithium jump networks and construct MSMs that remain Markovian across more than two orders of magnitude in the lag times used for their construction. The resulting MSMs faithfully reproduce mean-square displacements and recover rare diffusion processes that do not occur on AIMD timescales.
In addition to propagating lithium distributions, the MSM transition matrices provide mechanistic insight: their eigenvalues and eigenvectors encode characteristic relaxation timescales and dominant transport pathways.

Although demonstrated for defect-free crystalline Li$_x$Si$_y$ phases, the AIMD$\rightarrow$MLFF$\rightarrow$MSM framework is general and provides a transferable approach for describing lithium transport in amorphous materials, defect-mediated diffusion, and next-generation solid-state anodes.
\end{abstract}

\begin{tocentry}
\includegraphics[width=8.5cm]{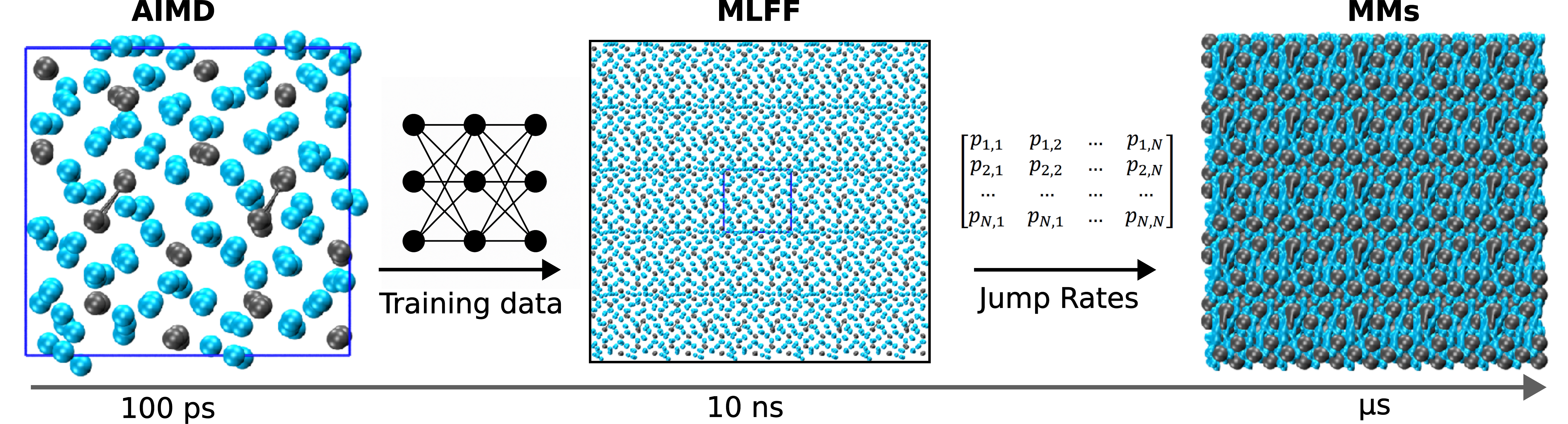}
\end{tocentry}

\section{Introduction}
Lithium-ion batteries power technologies ranging from portable electronics to electric vehicles and grid-scale storage, yet their performance and lifetime are tightly coupled to ion transport within electrode materials.~\cite{masias2021opportunities,zhu2021review,chen2020applications,duan2020building} In particular, the anode critically influences the achievable energy density, charging kinetics, and mechanical stability of the cell.~\cite{yang2018understanding,nzereogu2022anode,li2022fast,weiss2021fast} 
Graphite, the dominant commercial anode, provides reliable cycling but is constrained by a modest capacity of 372~mAh/g, motivating the search for higher-capacity alternatives.~\cite{zhao2024progress,asenbauer2020success,feng2018silicon} 
Silicon stands out as a promising candidate with a theoretical capacity of 3579~mAh/g in its fully lithiated Li$_{15}$Si$_4$ state, nearly ten times that of graphite.~\cite{ma2014si,zhang2021challenges} 

Despite this advantage, silicon anodes face severe challenges. Upon lithiation, they undergo volume changes of up to 380\%, leading to mechanical degradation and rapid capacity fading.~\cite{li2017research,zhang2021challenges,majeed2024silicon,Ma2015,Seo2025,Khan2024,Wang2023,Chen2019,Ibrahem2024} Understanding lithium transport is central to addressing these issues, as diffusion kinetics govern charge rates, stress evolution, and overall electrochemical stability. While amorphous silicon dominates in practical electrodes, its structural disorder complicates systematic analysis. Crystalline lithium silicides (Li$_x$Si$_y$), by contrast, provide well-defined diffusion pathways and serve as controlled model systems for probing lithium mobility and benchmarking computational methods.~\cite{dupke2012structural,zeilinger2013revision,kirsch2025li+, kirsch2026li+} Insights obtained from such ordered phases can be transferred to more complex amorphous systems, offering a clean and interpretable starting point for developing multiscale transport models.

Recent progress in machine learning has transformed our ability to model lithium diffusion at the atomic scale. Machine-learned force fields (MLFFs) have emerged as an efficient solution to the traditional accuracy--efficiency trade-off. By training directly on quantum-mechanical reference data, MLFFs can reach near-DFT accuracy while enabling nanosecond-scale simulations in large supercells. Frameworks such as Behler--Parrinello neural network potentials,~\cite{behler2007generalized} Gaussian approximation potentials (GAP),~\cite{bartok2010gaussian} and modern equivariant graph neural networks like NequIP~\cite{batzner20223} and MACE~\cite{batatia2022mace} have demonstrated remarkable versatility in capturing chemically diverse bonding environments. In the context of batteries, MLFFs have been successfully applied to lithium dynamics on metal surfaces,~\cite{tian2023deep} in amorphous electrodes,~\cite{deringer2020modelling} in graphite,~\cite{cheng2025large} and in solid electrolytes such as Li$_3$TiCl$_6$,~\cite{selvaraj2024exploring} highlighting their potential for a broad range of electrochemical systems.

The development of reactive and machine-learned force fields for lithium diffusion in silicon anodes is an active area of research.~\cite{Xu2020,Fu2023,Fu2024,OlououGuifo2024,Chen2024} Although these approaches dramatically extend the accessible time and length scales compared to \textit{ab initio} molecular dynamics, they remain fundamentally atomistic. Yet real anodes exhibit mesoscale complexity: nanostructured morphologies, heterogeneous lithium concentrations, and evolving microstructures designed to accommodate large volume expansion during lithiation.~\cite{Ma2015,Seo2025,Khan2024,Wang2023,Chen2019,Ibrahem2024} In such environments, lithium motion involves collective and correlated ion dynamics spanning microseconds and micrometers. Bridging this gap between atomic detail and mesoscale transport requires systematic coarse-graining strategies that preserve the underlying physics of atomistic motion while remaining quantitatively consistent with first-principles energetics.

A particularly powerful strategy, which has so far seen limited application to solid-state Li transport, is to extract lithium jump statistics from long MLFF trajectories and represent them using Markov state models (MSMs).~\cite{Husic2018,Prinz2011,Chodera2014}
MSMs recast ion motion as probabilistic transitions between (meta)stable sites, providing a rigorous statistical framework for coarse-graining atomistic dynamics and extrapolating diffusion behavior over extended temporal and spatial scales. In this representation, the long-timescale evolution of the system is captured by the repeated application of a transition matrix, allowing atomistic dynamics to be projected onto a coarse-grained, stochastic model. A key advantage of MSMs is that they make otherwise inaccessible timescales tractable through efficient propagation of state probabilities, i.e., several picoseconds of system dynamics can be recovered through simple matrix--vector multiplications. 

In the field of molecular dynamics, Markov state models (MSMs) have historically been
applied predominantly to classical biomolecular simulations, including protein and peptide
folding, ligand binding, and nucleic-acid conformational transitions
\cite{Prinz2011,Husic2018,Chodera2014}. In contrast, MSMs constructed directly from
atomistic MD trajectories for inorganic crystalline materials remain scarce.\cite{dressler2020dynamical, dressler2020exploring} Studies of ionic
transport in solids have instead traditionally relied on lattice-based jump-diffusion formalisms
or kinetic Monte Carlo simulations parameterized by transition-state theory or nudged
elastic band migration barriers \cite{Deng2022, Canepa2023, Deng2023, VanderVen2001, Gustafsson2025, dressler2016proton, hanseroth2025hydroxide}. These methods have long served as the standard framework for
describing thermally activated hopping processes in periodic lattices, whereas MSM-like
coarse-graining directly from MD trajectories is only beginning to emerge for solid-state ion
transport. 

The reliability of an MSM depends critically on two factors: the choice of lag time and the quality of sampling. The lag time must be long enough to average out rapid local fluctuations, ensuring Markovian behavior, but also short enough to resolve the relevant kinetic processes; if chosen too large, distinct transition pathways may be lumped together and important dynamical information lost. At the same time, sufficient statistics are needed to estimate transition probabilities with confidence. These conditions are particularly challenging for \textit{ab initio} molecular dynamics, where accessible trajectories are typically short. Validation procedures such as the Chapman--Kolmogorov test and the analysis of implied timescales provide practical means to assess whether an MSM is consistent and predictive.

In this work, we present a unified bottom-up workflow, AIMD~$\rightarrow$~MLFF~$\rightarrow$~MSM, that links quantum-accurate barriers to mesoscale transport. AIMD supplies high-fidelity reference data for model training, MLFFs extend simulation times and system sizes to capture rare hopping events, and MSMs coarse-grain the resulting trajectories into statistically consistent transport models. We demonstrate this approach for crystalline Li--Si phases as a controlled test case, establishing a quantitative and transferable foundation for modeling lithium transport in amorphous and defect-rich silicon anodes relevant to next-generation battery technologies.

\section{Theory}

\subsection{Markov State Models}

To construct a Markov state model (MSM) from molecular dynamics (MD) trajectories, the continuous phase space is discretized into a finite set of states. Here, we adopt a single-particle description and treat the motion of each Li ion as a Markov process on a lattice of crystallographic sites. The dynamics are then described by a state probability vector $x^t$, whose components give the probabilities of finding a given Li ion at each site at time $t$. This probability vector evolves under the action of a transition matrix $\mathcal{M}^\tau$ defined at lag time $\tau$:
\begin{equation}
x^{t+\tau} = \mathcal{M}^\tau x^t.
\label{eq:prop}
\end{equation}

The transition matrix $\mathcal{M}^\tau$ is estimated by counting transitions between discretized states within intervals of length $\tau$ along the MD trajectory.\cite{Prinz2011,Bowman2014} The discretization of phase space introduces a minimal lag time: if $\tau$ is too short, rapid local fluctuations (``rattling'') dominate, and the resulting dynamics deviate from the Markov property.\cite{Prinz2011,Husic2018} Reliable statistics further require a large number of independent intervals of length $\tau$, which is a practical limitation for AIMD simulations.

The Chapman--Kolmogorov test provides a direct assessment of Markovianity by comparing transition matrices sampled at multiples of the lag time with powers of the single-step transition matrix,\cite{Prinz2011,Noe2008}
\begin{equation}
\left(\mathcal{M}^\tau_{\rm sampled}\right)^n \approx \mathcal{M}^{n\tau}_{\rm sampled}.
\label{eq:chap}
\end{equation}

In addition, the spectral decomposition of $\mathcal{M}^\tau$ provides mechanistic insight.\cite{Noe2008,Chodera2014,Sittel2016} The leading eigenvalue $\lambda_1 = 1$ corresponds to the stationary distribution of the system, while all other eigenvalues satisfy $|\lambda_k| < 1$. Their associated eigenvectors describe relaxation processes from non-equilibrium states, with eigenvalues close to one corresponding to slow dynamics and smaller eigenvalues to faster processes. For each nontrivial mode $k \geq 2$, the implied timescale is defined as
\begin{equation}
t_k(\tau) = -\frac{\tau}{\ln \lambda_k(\tau)}.
\label{eq:tau_conv}
\end{equation}

The appearance of plateaus in implied timescales as a function of $\tau$ indicates the characteristic timescales of the underlying processes and defines a lower bound on lag times for which a consistent MSM can be constructed.\cite{Prinz2011,Husic2018,Chodera2014,Bowman2014,Noe2008}.

\subsection{Central Idea of the AIMD~$\rightarrow$~MLFF~$\rightarrow$~MSM Approach}

This work integrates \textit{ab initio} molecular dynamics, machine-learned force fields, and Markov state models into a unified multiscale framework. AIMD is employed to generate reference datasets for training and validating MLFFs. The fine-tuned MLFFs then enable nanosecond-scale molecular dynamics trajectories, from which lithium jump statistics are extracted. These statistics are subsequently coarse-grained into MSMs, providing a stochastic representation of ion transport over extended spatial and temporal scales. The following sections detail the simulation protocols and model construction steps.

\subsubsection{Ab Initio Molecular Dynamics Simulations}

The \textit{ab initio} molecular dynamics trajectories used in this study were taken from our previously published work,~\cite{kirsch2022atomistic} in which defect-free Li$_{12}$Si$_7$ and Li$_{13}$Si$_4$ systems were simulated using the CP2K package.~\cite{vandevondele2005quickstep} Simulations were performed in the NVT ensemble using the Becke--Lee--Yang--Parr (BLYP) exchange--correlation functional,~\cite{Lee1988,Becke1988} GTH pseudopotentials,~\cite{goedecker1996separable} and DZVP-MOLOPT-SR-GTH basis sets.~\cite{vandevondele2007gaussian} Dispersion interactions were included via the DFT-D3 method,~\cite{grimme2010consistent} and a time step of 0.5~fs was employed together with a Nosé--Hoover chain thermostat.~\cite{martyna1992nose} Both systems were equilibrated and simulated at 500~K, producing 100~ps trajectories. From each trajectory, we uniformly extracted either 200 or 2000 frames to construct the training datasets, ensuring diverse coverage of local atomic environments across the full simulation length.

\subsubsection{Fine-Tuning of MLFFs}

Machine-learned force fields were developed within the MACE framework,~\cite{batatia2022mace} implemented via the MACE Python package (v0.3.10). We initialized from the publicly available MACE-MP-0 foundation model and fine-tuned it on system-specific DFT reference data extracted from the AIMD trajectories.~\cite{batatia2025foundationmodelatomisticmaterials,mptrj}
For each system, two fine-tuned models were trained: one on a reduced dataset (200 frames) and one on an extended dataset (2000 frames). A detailed explanation of the fine-tuning protocol is given in the Supporting Information. The fine-tuning procedure was carried out using the workflow we have already applied to other systems\cite{grunert2025modeling, hanseroth2025optimizing, flototto2025large} and recently implemented in the \texttt{aMACEing\_toolkit} package.~\cite{haenseroth2025amaceingtoolkit}

\subsubsection{Molecular Dynamics Simulations and Validation of MLFFs}

Molecular dynamics simulations were performed using the fine-tuned MACE models interfaced through the Atomic Simulation Environment (ASE).~\cite{Larsen2017ASE} All simulations were carried out in the NVT ensemble with a Langevin thermostat to maintain a constant temperature of 500~K. To validate the results, equivalent simulations were repeated with the LAMMPS package using a Nosé--Hoover chain thermostat.~\cite{LAMMPS} An integration time step of 0.5~fs was used throughout. For each Li$_x$Si$_y$ system, trajectories of up to 30~ns were generated. These extended timescales were crucial for capturing rare hopping events and collective ionic motion, and for achieving statistically converged transport properties. Periodic boundary conditions were applied in all three spatial directions to minimize finite-size effects. The molecular dynamics simulations were prepared using the workflow provided by the \texttt{aMACEing\_toolkit} package.~\cite{haenseroth2025amaceingtoolkit}

\subsubsection{Markov Model Construction and Convergence}
\label{sec:create_markov}

The construction of Markov state models from MD trajectories consists of four steps:

\textbf{Step 1: Discretization of phase space.}

Lithium transport in Li$_x$Si$_y$ compounds can be represented as a sequence of hops between lattice sites. In the crystalline phases studied here, we identify these states directly from the crystallographic lithium positions. In amorphous Li--Si, however, no predefined lattice exists; states would need to be obtained, for example, by geometric clustering of lithium positions or by locating local minima in the potential energy landscape. Although we do not construct MSMs for amorphous systems in the present work, the rigorous protocol established here for crystalline Li$_x$Si$_y$, in particular the determination of minimal lag times, sampling requirements, and validation through implied timescales and Chapman--Kolmogorov tests, provides essential guidelines for extending MSM-based transport modeling to amorphous phases where state definitions are less straightforward but lithium hopping still gives rise to Markovian long-time dynamics.

If the simulated system contains $N$ lithium atoms, the system is represented by an $N$-dimensional occupation vector $\vec{x}$, where the $i$th component corresponds to the fraction of lithium atoms residing at the $i$th lattice site.

\textbf{Step 2: Sampling of the transition matrix.}
Let $\mathcal{S} = \{1, \dots, N\}$ denote the set of sites and $s(t) \in \mathcal{S}$ the site occupied at time $t$. Sampling at a lag time $\tau$ yields pairs $\big(s(t), s(t+\tau)\big)$, from which the count matrix is obtained as
\begin{equation}
C_{ij}(\tau) = \#\{t : s(t) = j,\; s(t+\tau) = i\}.
\end{equation}
The corresponding column-stochastic transition matrix is
\begin{equation}
\mathcal{M}^\tau_{ij} =
\frac{C_{ij}(\tau)}{\sum_{k} C_{kj}(\tau)}, 
\qquad
\sum_i \mathcal{M}^\tau_{ij} = 1.
\label{eq:count}
\end{equation}

\textbf{Step 3: Validation of Markovianity.}
The validity of the Markov assumption was assessed through (i) the analysis of implied timescales and (ii) the Chapman--Kolmogorov (CK) test.

The plateau region of the implied timescales [eq.~\ref{eq:tau_conv}] defines the minimal lag time $\tau_\mathrm{min}$ for which the dynamics can be approximated as Markovian.
Choosing a lag time $\tau$ from within the plateau region of these modes ensures that the resulting MSM is both Markovian and predictive.

The CK test provides an independent assessment by comparing transition matrices sampled at multiples of the lag time $\tau$ with the corresponding powers of the single-step transition matrix. The relative deviation is quantified as
\begin{equation}
\mathrm{err}(n) =
\frac{\left\| \mathcal{M}_{\mathrm{sampled}}^{\,n\tau}
      - \left(\mathcal{M}^\tau\right)^n \right\|_2}
     {\left\| \mathcal{M}_{\mathrm{sampled}}^{\,n\tau} \right\|_2},
\label{eq:err}
\end{equation}
where $\|\cdot\|_2$ denotes the matrix 2-norm and $n$ is the length of the Markov chain. Small values of $\mathrm{err}(n)$ indicate that the dynamics sampled directly from the MD trajectory are consistent with those predicted by the Markov model.

\textbf{Step 4: Calculation of the mean-square displacement (MSD).}
Transport properties are reconstructed by propagating displacements through the MSM. The matrix element $\left( \mathcal{M}^{n\tau} \right)_{ij}$ gives the probability of transfer from site $j$ to site $i$ after $n$ steps. Denoting the distance between sites $i$ and $j$ as $d_{ij}$, the mean-square displacement is given by
\begin{equation}
\mathrm{MSD}(n\tau) = \sum_{ij} \left( \mathcal{M}^{n\tau} \right)_{ij} d_{ij}^{2}.
\label{eq:msd}
\end{equation}
The diffusivity $D$ is then obtained from the long-time slope of the MSD using the Einstein relation:
\begin{equation}
\label{eq:einstein}
D = \lim_{t\to\infty}\frac{1}{2d}\,\frac{\mathrm{d}}{\mathrm{d}t}\,\langle |\Delta \mathbf{r}(t)|^2 \rangle,
\qquad d=3.
\end{equation}

Because $d_{ij}$ in Eq.~\ref{eq:msd} cannot exceed half the length of the simulation cell under periodic boundary conditions, diffusion is effectively confined to the dimensions of the MD supercell. Consequently, the MSD computed from the transition matrices saturates and exhibits a plateau at very long times.
To mitigate this artifact, we construct transition matrices for replicated supercells, thereby shifting the plateau to timescales beyond 1~ns and recovering the diffusive regime within the accessible temporal window.

\bigskip

\begin{figure}[htbp]
    \centering
    \begin{minipage}{0.40\textwidth}
        \centering
        \includegraphics[width=\textwidth]{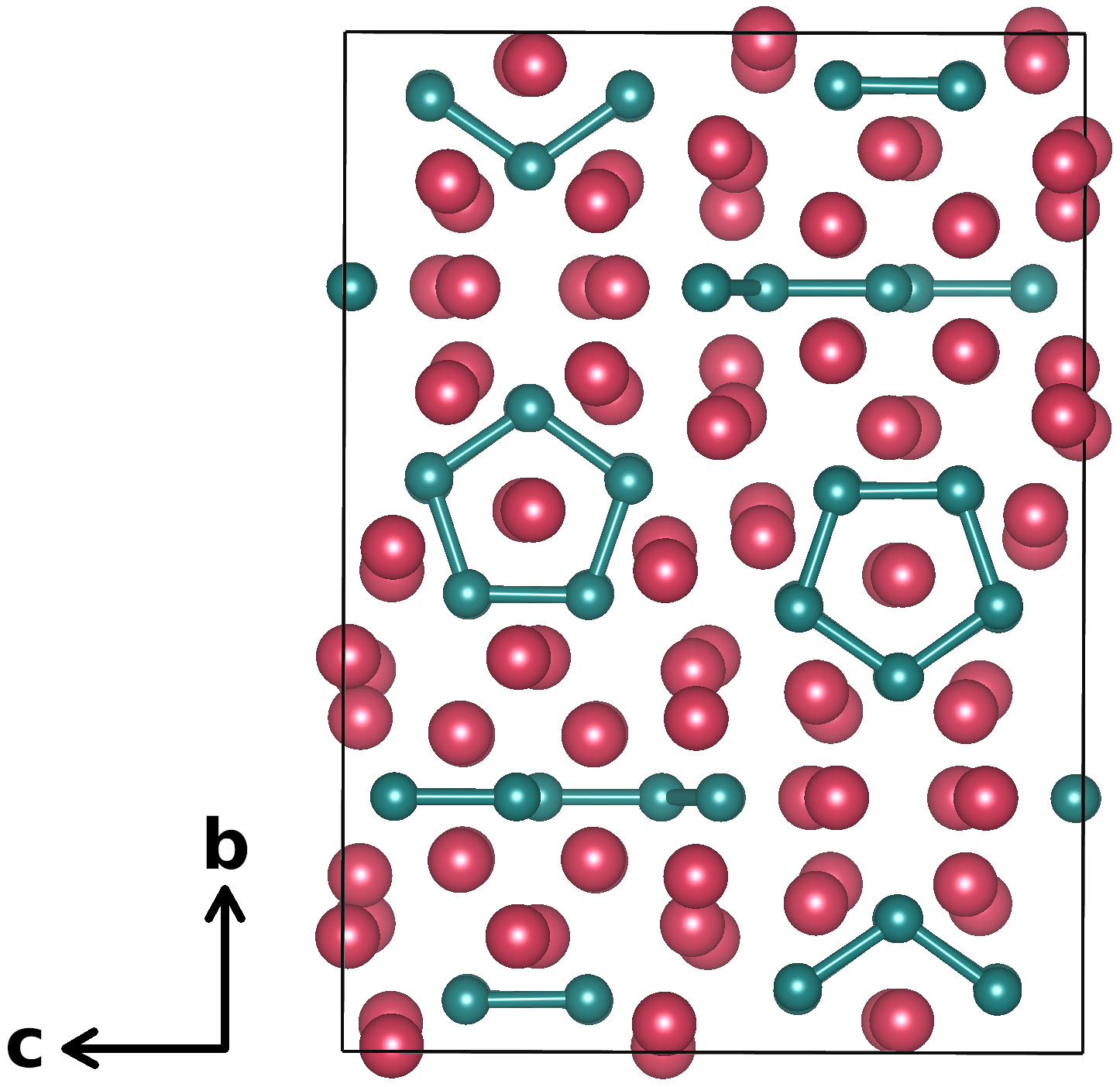}
        \put(-150,200){\makebox(0,0)[lt]{\textbf{(a)}}}
    \end{minipage}
    \hfill
    \begin{minipage}{0.49\textwidth}
        \centering
        \includegraphics[width=\textwidth]{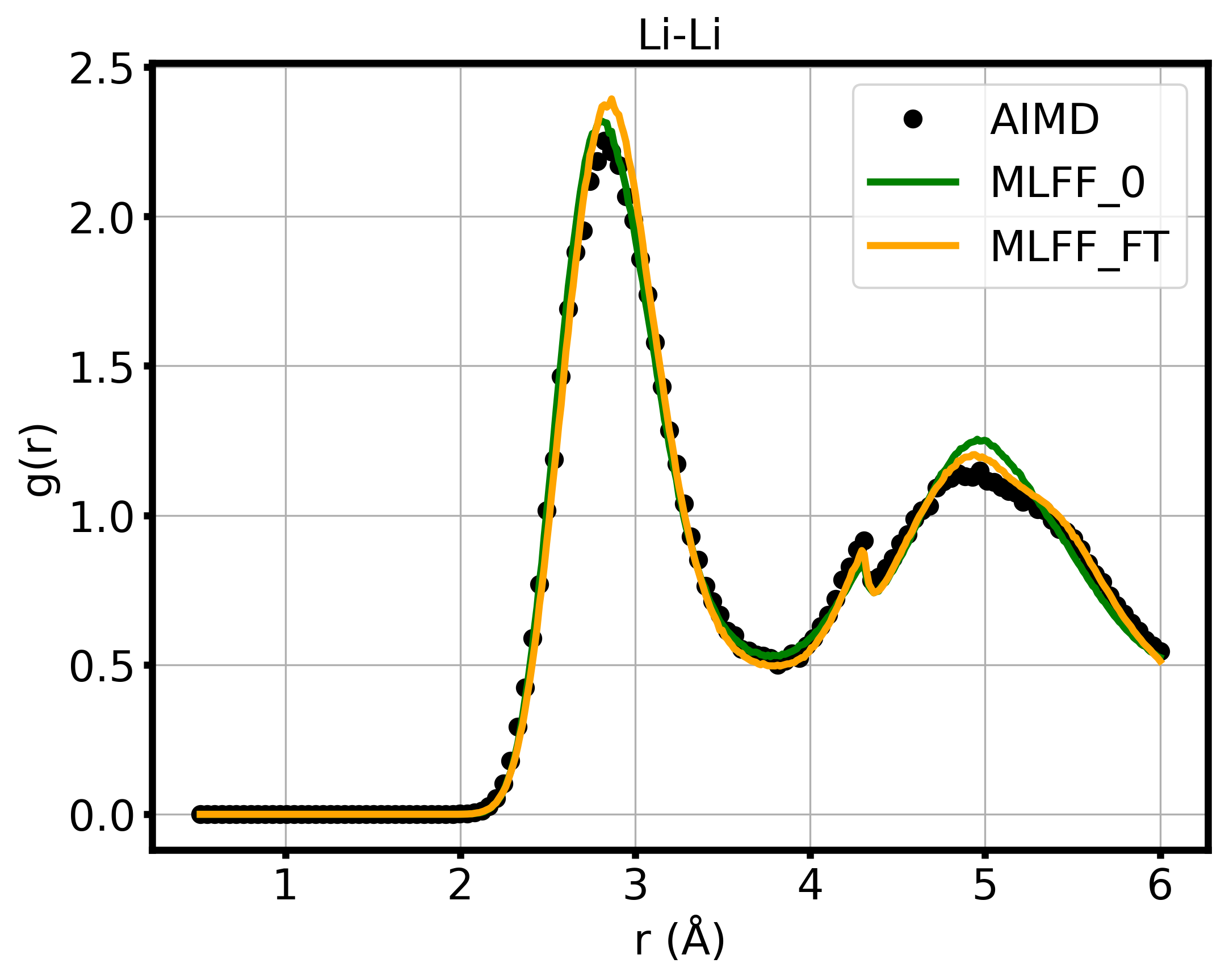}
        \put(-200,200){\makebox(0,0)[lt]{\textbf{(b)}}}
    \end{minipage}
    
\vspace{0.3cm}

    \begin{minipage}{0.49\textwidth}
        \centering
        \includegraphics[width=\textwidth]{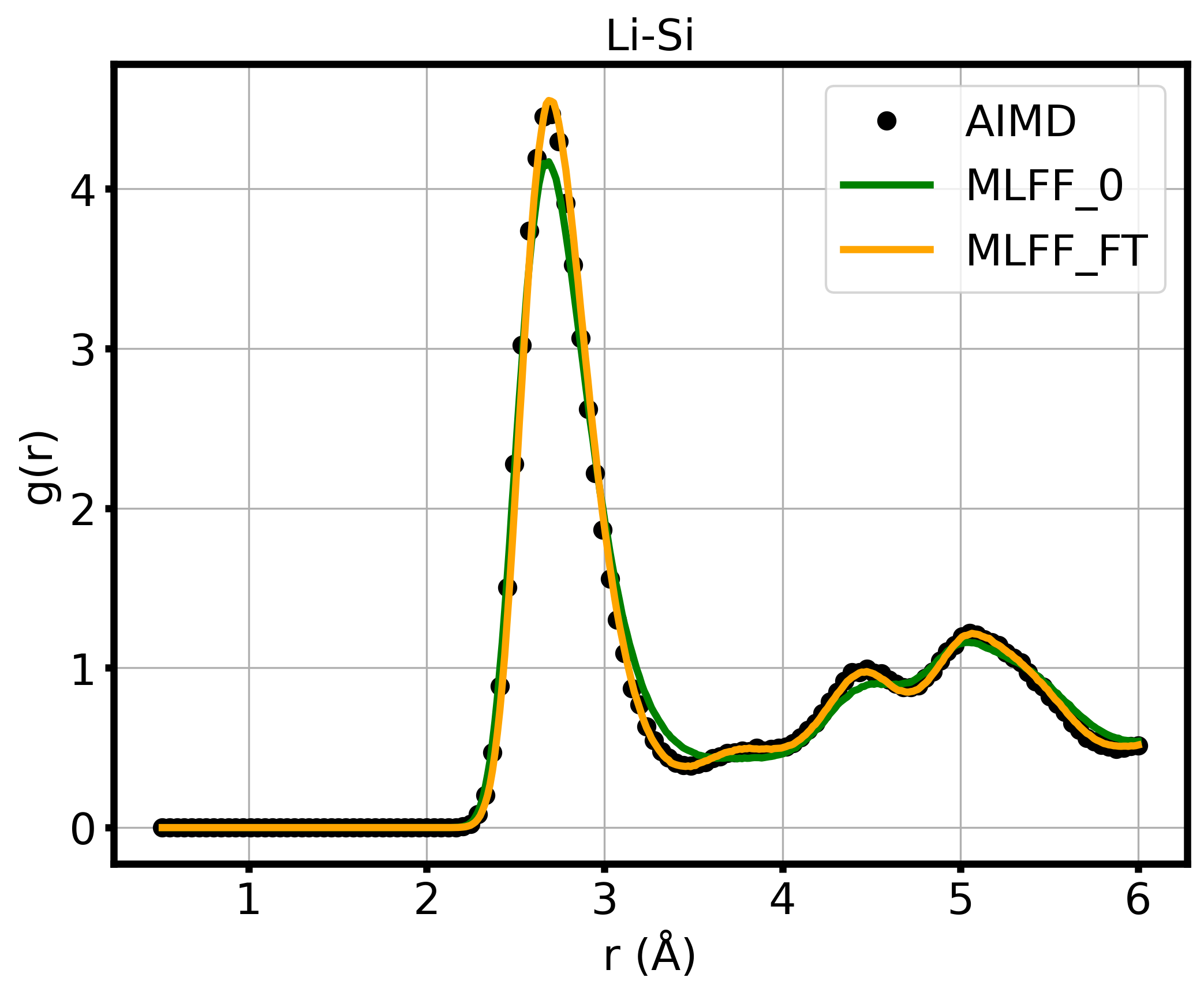}
        \put(-200,200){\makebox(0,0)[lt]{\textbf{(c)}}}
    \end{minipage}
    \hfill
    \begin{minipage}{0.49\textwidth}
        \centering
        \includegraphics[width=\textwidth]{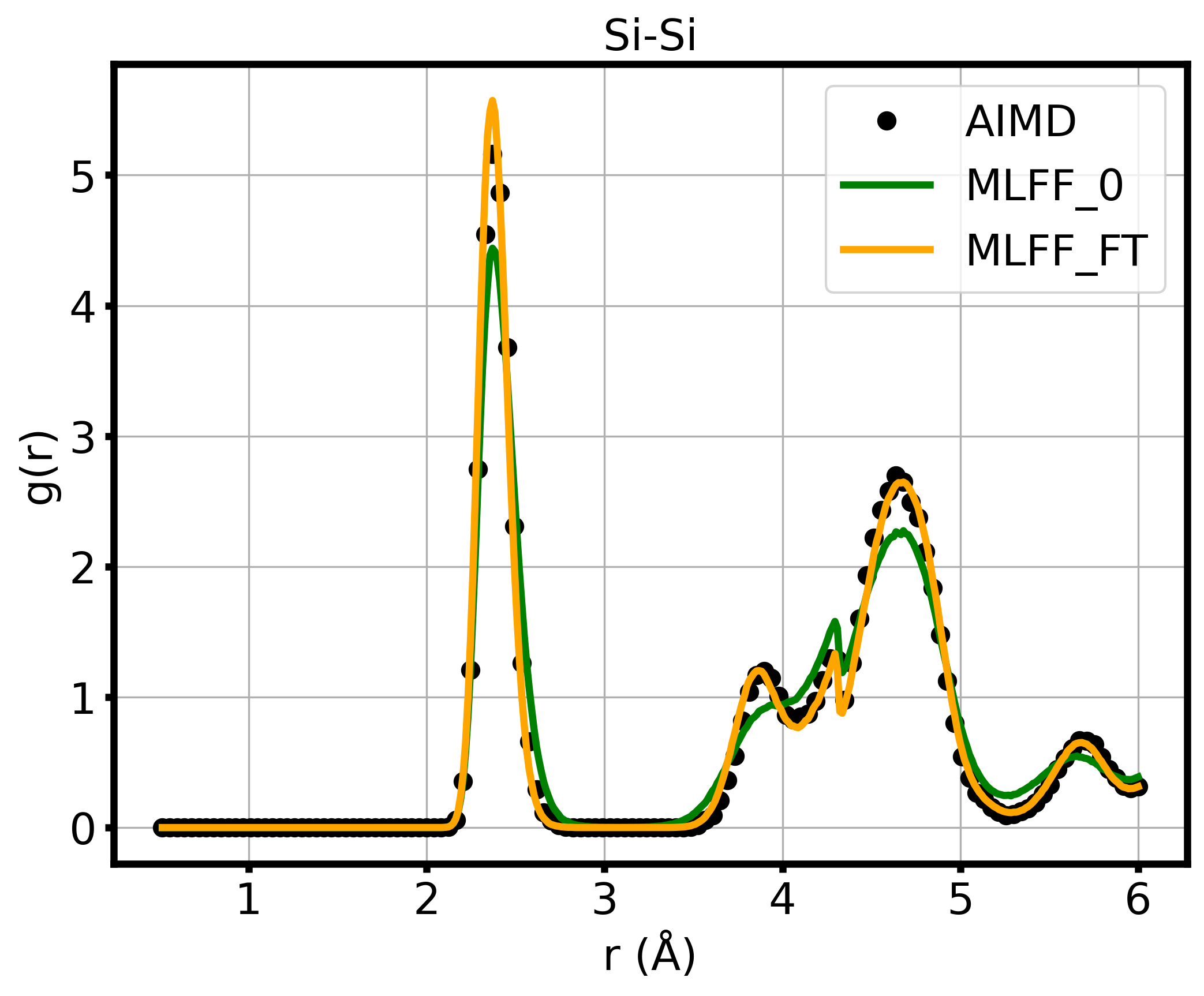}
        \put(-200,200){\makebox(0,0)[lt]{\textbf{(d)}}}
    \end{minipage}

    \caption{(a) Snapshot of the Li$_{12}$Si$_7$ crystal structure. Lithium and silicon atoms are shown in red and teal, respectively. Panels (b)--(d) illustrate comparisons of the radial distribution function $g(r)$ obtained from AIMD (black filled circles), the MACE foundation model (green line), and the fine-tuned model (orange line) for Li--Li, Li--Si, and Si--Si pairs, respectively.}
    \label{rdf}
\end{figure}

\begin{figure}[htbp]
    \centering
    \begin{minipage}{0.49\textwidth}
        \centering
        \includegraphics[width=0.6\textwidth]{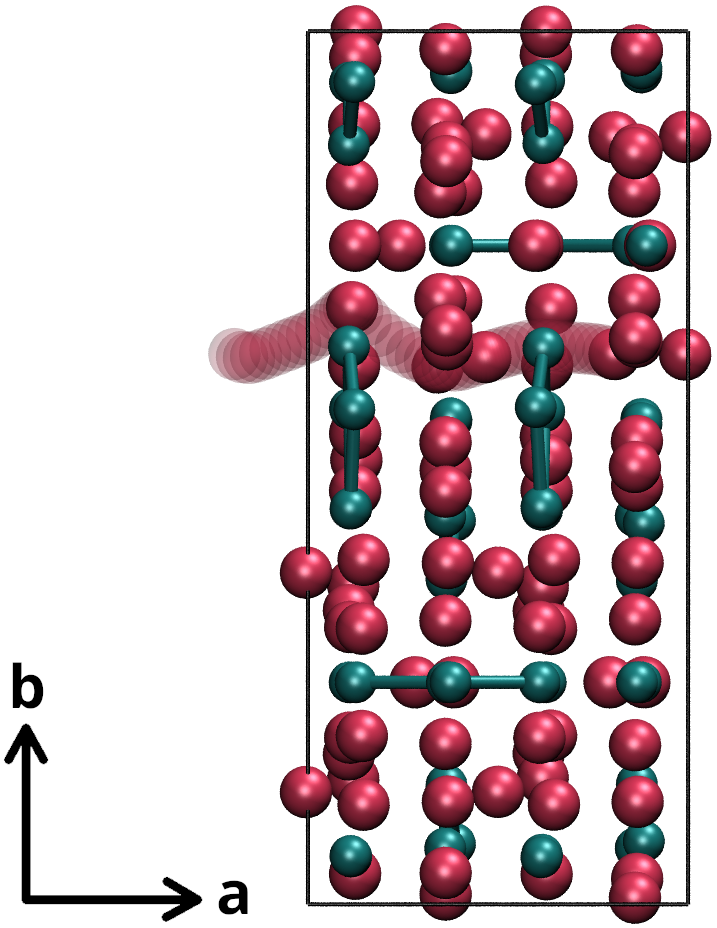}
        \put(-150,190){\makebox(0,0)[lt]{\textbf{(a)}}}
    \end{minipage}
    \hfill
    \begin{minipage}{0.49\textwidth}
        \centering
        \includegraphics[width=\textwidth]{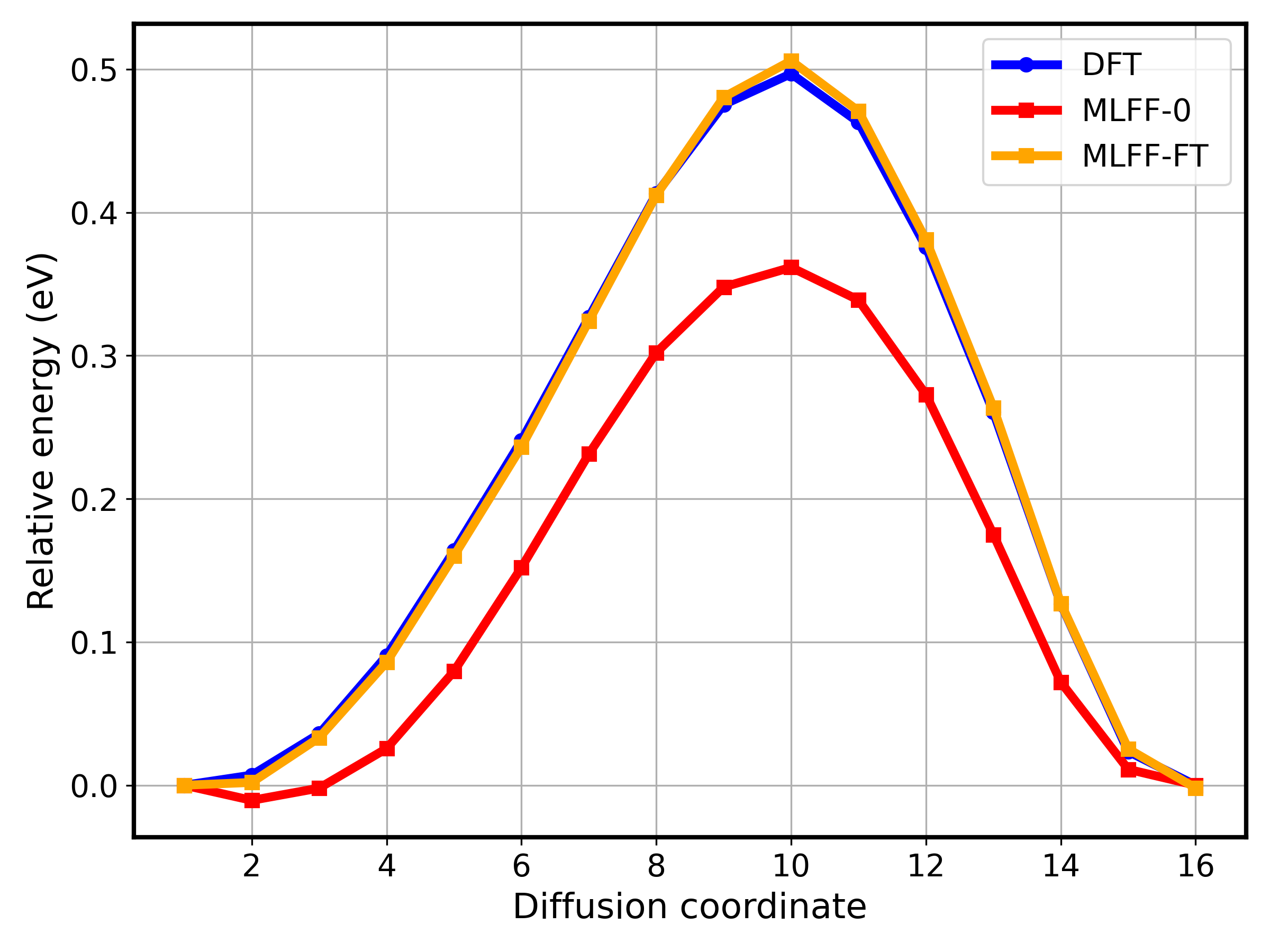}
        \put(-200,190){\makebox(0,0)[lt]{\textbf{(b)}}}
    \end{minipage}

\vspace{0.3cm}

    \begin{minipage}{0.49\textwidth}
        \centering
        \includegraphics[width=\textwidth]{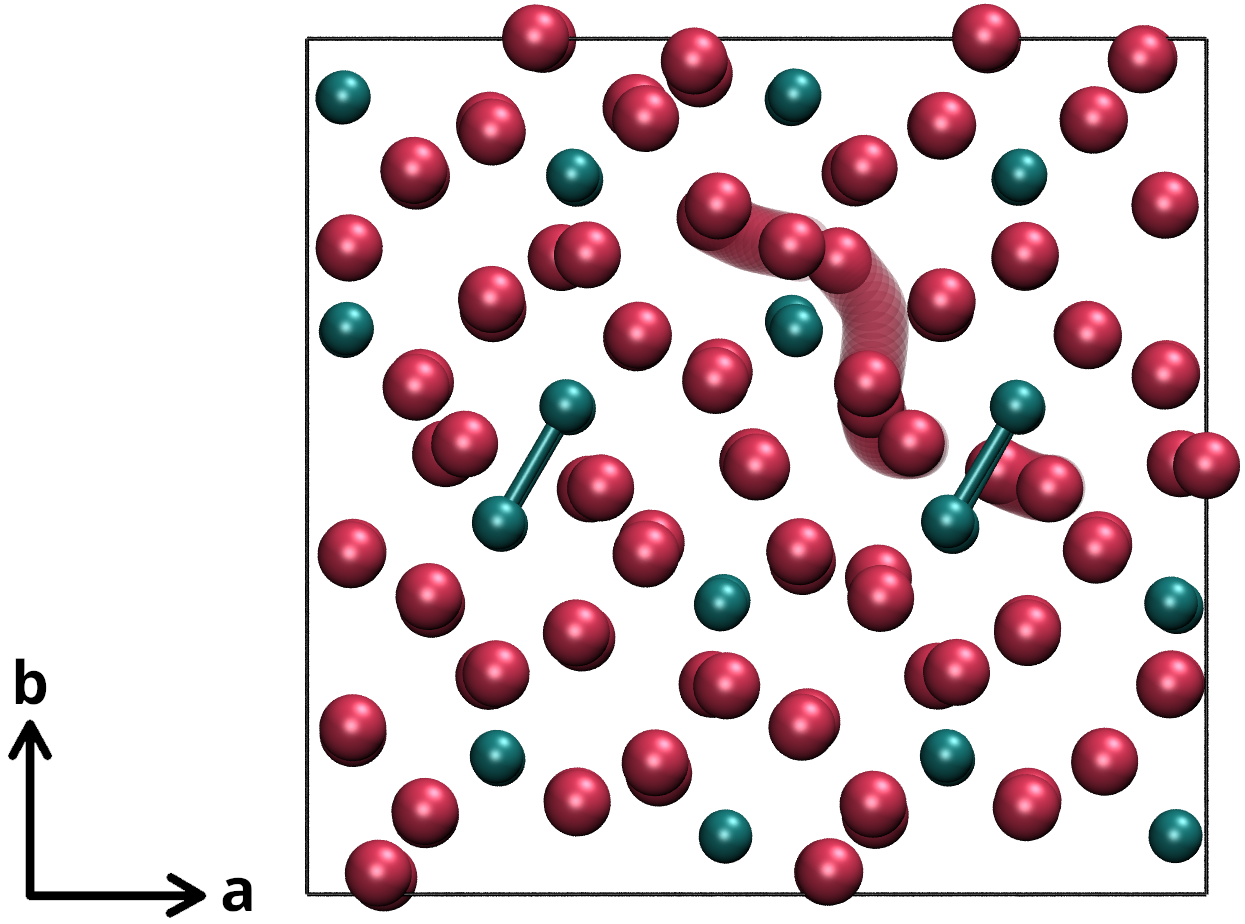}
        \put(-200,190){\makebox(0,0)[lt]{\textbf{(c)}}}
    \end{minipage}
    \hfill
    \begin{minipage}{0.49\textwidth}
        \centering
        \includegraphics[width=\textwidth]{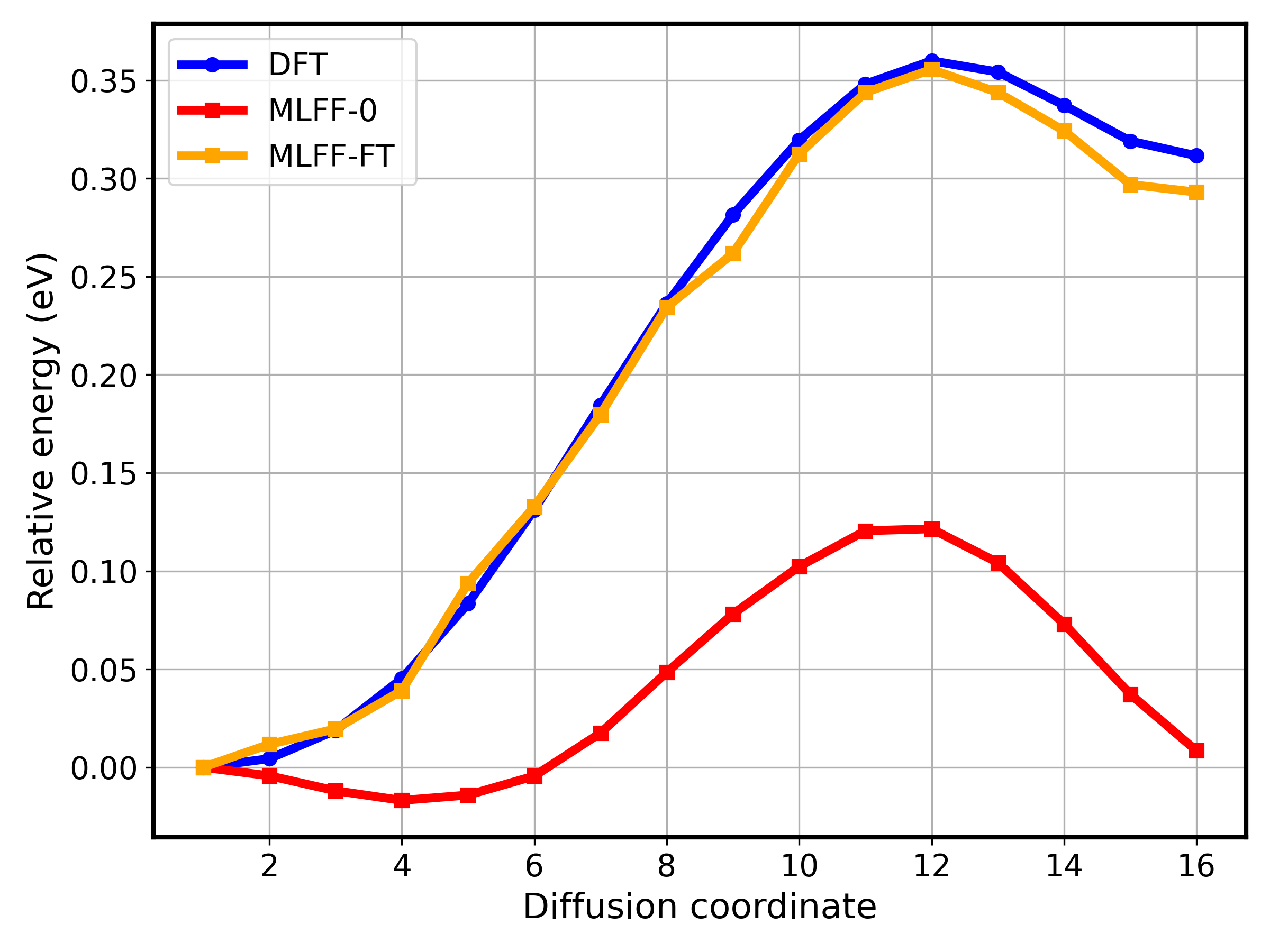}
        \put(-200,190){\makebox(0,0)[lt]{\textbf{(d)}}}
    \end{minipage}

    \caption{Representative lithium migration pathways and activation energies. Panels (a) and (c) show selected migration paths in Li$_{12}$Si$_7$ and Li$_{13}$Si$_4$, respectively. Panels (b) and (d) show corresponding NEB energy profiles computed using DFT (blue), the MACE foundation model (red), and the fine-tuned model (orange).}
    \label{neb}
\end{figure}

\section{Results}

We began by assessing the accuracy of the machine-learned force fields trained within the MACE framework. Training datasets were generated for Li$_{12}$Si$_7$ and Li$_{13}$Si$_4$, and training performance was evaluated using root-mean-square errors (RMSEs) in total energies and atomic forces (Table~S1). Across both phases, the fine-tuned models achieved energy errors below 2~meV~atom$^{-1}$ and force errors below 30~meV~\AA$^{-1}$. Additional validation against direct DFT evaluations along independent MLFF-driven MD trajectories (Table~S2) showed comparable deviations in forces and energies, suggesting that the models can reproduce short-time dynamical behavior such as lithium hopping. The resulting MLFFs therefore provide a reliable description of the potential-energy surfaces for extended simulations of lithium transport in Li$_x$Si$_y$ compounds. 

Previous studies have also reported machine-learned force fields for lithium silicides.~\cite{Fu2023,Fu2024,Xu2020} These models, typically based on deep neural network architectures, reported force RMSEs on the order of 100~meV~\AA$^{-1}$. In comparison, our fine-tuned MACE models achieve lower errors with similar training-set sizes, reflecting the advantages of equivariant graph neural networks that preserve rotational symmetries and incorporate angular and many-body correlations in local atomic environments. These features contribute to improved data efficiency and a more accurate representation of local structural interactions within the tested systems. 

We next examined whether the fine-tuned models also reproduce structural correlations characteristic of the underlying phases. Figure~\ref{rdf} compares radial distribution functions (RDFs) of Li$_{12}$Si$_7$ at 500~K obtained from AIMD, the pretrained MACE foundation model, and the fine-tuned MLFF. While the foundation model captures short-range features of $g(r)$, systematic deviations appear at intermediate and long distances, particularly for Si--Si correlations, reflecting the challenge of describing diverse bonding motifs in Li--Si networks with a general-purpose model. Fine-tuning substantially improves agreement across all pair types: the positions and amplitudes of Li--Si and Si--Si peaks are accurately reproduced, and the long-range decay of $g(r)$ closely follows AIMD. The Li--Li peak amplitude is slightly overestimated, but the overall structural ordering remains consistent with DFT-based reference data. A similar level of agreement is observed for Li$_{13}$Si$_4$ (Fig.~S1), confirming that the fine-tuned MLFFs reliably capture both local and extended order across distinct crystalline Li--Si compositions. 

Building on this structural fidelity, we further evaluated kinetic accuracy by comparing lithium migration barriers computed with the nudged elastic band (NEB) method.~\cite{Henkelman1999} Diffusion pathways identified from AIMD trajectories were relaxed at the DFT level and then recomputed using both the foundation and fine-tuned MLFFs. Figure~\ref{neb} shows representative migration paths and their corresponding energy profiles for Li$_{12}$Si$_7$ and Li$_{13}$Si$_4$. The foundation model systematically underestimates and occasionally overestimates the migration barriers, reflecting its emphasis on broad transferability at the expense of localized precision. In contrast, the fine-tuned MLFFs reproduce DFT-calculated barriers within 2--5\% error, in some cases yielding nearly indistinguishable profiles (see the Supporting Information). In Li$_{13}$Si$_4$, a dominant one-dimensional diffusion channel was identified along a crystallographic axis, with its barrier precisely captured by the fine-tuned model (Fig.~\ref{fig:jumps_six}). Accurate barriers are essential because diffusivities follow Arrhenius-type scaling, where even modest deviations propagate exponentially. By matching DFT-calculated barriers across multiple migration paths in both Li$_{12}$Si$_7$ and Li$_{13}$Si$_4$, the fine-tuned MLFFs establish a reliable foundation for transport modeling. 

This accuracy ensures that lithium jump statistics extracted from long MLFF simulations can be coarse-grained into MSMs without systematic bias in the underlying energetics. The accurate reproduction of both structural correlations (Fig.~\ref{rdf}) and migration barriers (Fig.~\ref{neb}) establishes that our fine-tuned MLFFs capture the essential lithium--silicon interactions that govern local coordination. In amorphous Li--Si systems, these same local environments, characterized by specific Li--Si coordination numbers and bond lengths, dominate the potential energy landscape. The demonstrated accuracy of our MLFFs in describing these fundamental interactions provides confidence in their transferability to amorphous phases, where the primary difference lies in the long-range structural disorder rather than the local bonding chemistry. 

\FloatBarrier

\subsection{Lithium Diffusion and Finite-Size Effects}

\begin{figure}[htbp]
  \centering
  \begin{minipage}{0.49\textwidth}
    \centering
    \includegraphics[width=\textwidth]{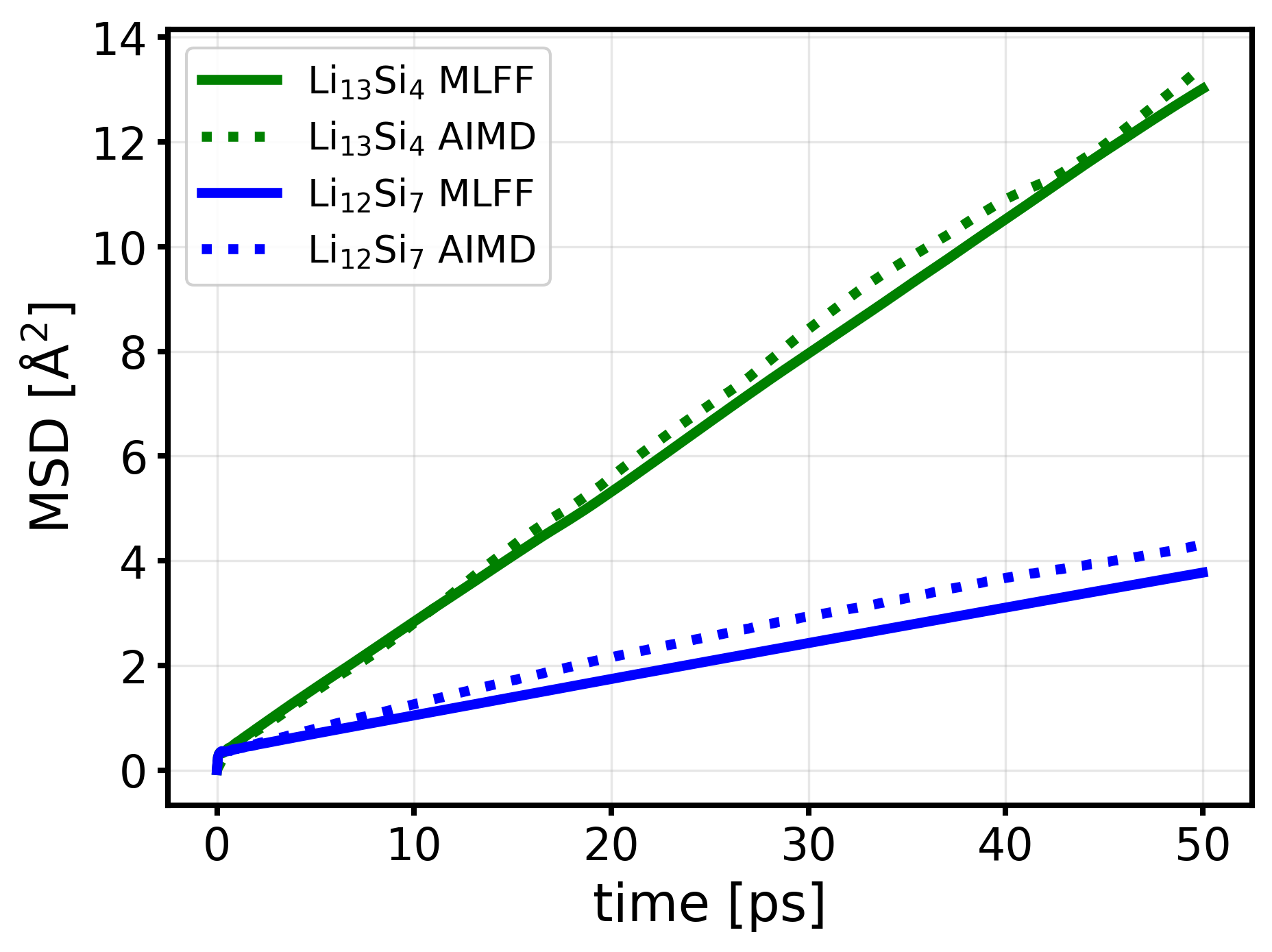}\\[-0.2em]
    \small (a) 
  \end{minipage}
  \begin{minipage}{0.49\textwidth}
    \centering
    \includegraphics[width=\textwidth]{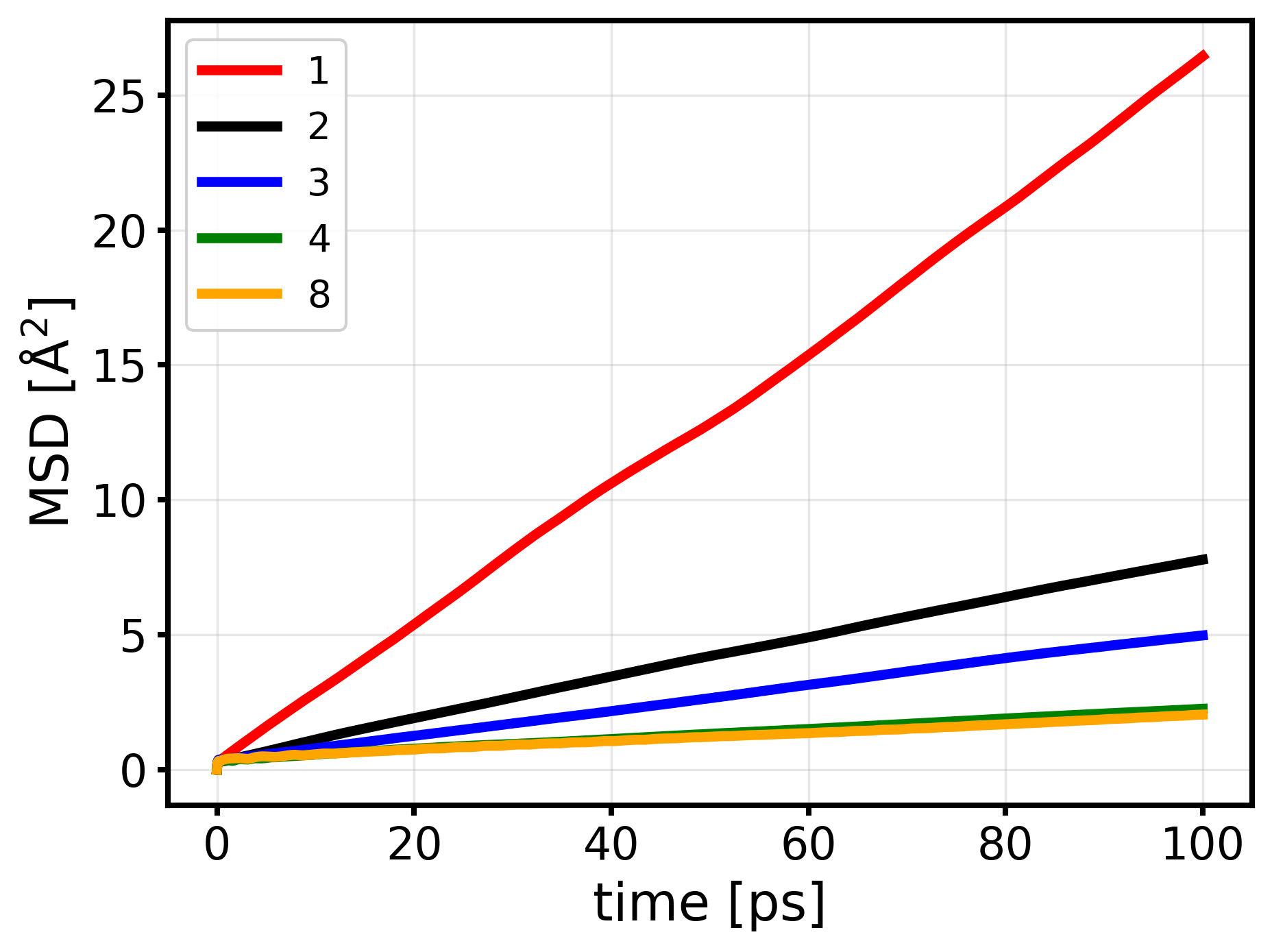}\\[-0.2em]
    \small (b)  
  \end{minipage}
  \caption{(a) Mean-square displacement (MSD) from AIMD and MLFF simulations. (b) Convergence of the diffusion coefficient (MSD slope) with respect to system size in the $x$-direction from MLFF simulations. Values 1, 2, 3, 4, and 8 in the legend correspond to a 1-, 2-, 3-, 4-, and 8-fold increase of the supercell dimension in the $x$-direction. Apparent diffusivities computed in AIMD-accessible supercells exhibit a pronounced finite-size bias, whereas larger MLFF supercells approach the asymptotic, size-independent limit.}
  \label{fig:msd_aimd_mlff}
\end{figure}

Having established that the fine-tuned MLFFs reproduce both static energetics and migration barriers with near-DFT accuracy, we next employ them to explore lithium diffusion dynamics over extended timescales and to quantify finite-size effects on the computed transport coefficients. The mean-square displacements for Li$_{12}$Si$_7$ and Li$_{13}$Si$_4$ obtained from MLFF and AIMD simulations are shown in Fig.~\ref{fig:msd_aimd_mlff}a. MSDs from both methods are in very good agreement for both compounds, further confirming the accuracy of the MLFFs. 

Li$_{12}$Si$_7$ and Li$_{13}$Si$_4$ were selected because, compared to other crystalline lithium silicides such as Li$_{15}$Si$_4$, LiSi, and Li$_{17}$Si$_4$, they exhibit exceptionally low migration barriers for lithium diffusion even in defect-free systems. As a result, converged diffusion coefficients can be obtained on AIMD timescales, allowing the MSDs of these compounds to serve as meaningful benchmarks for evaluating the accuracy of the MLFFs. 

\begin{figure}[htbp]
  \centering
    \includegraphics[width=0.6\linewidth]{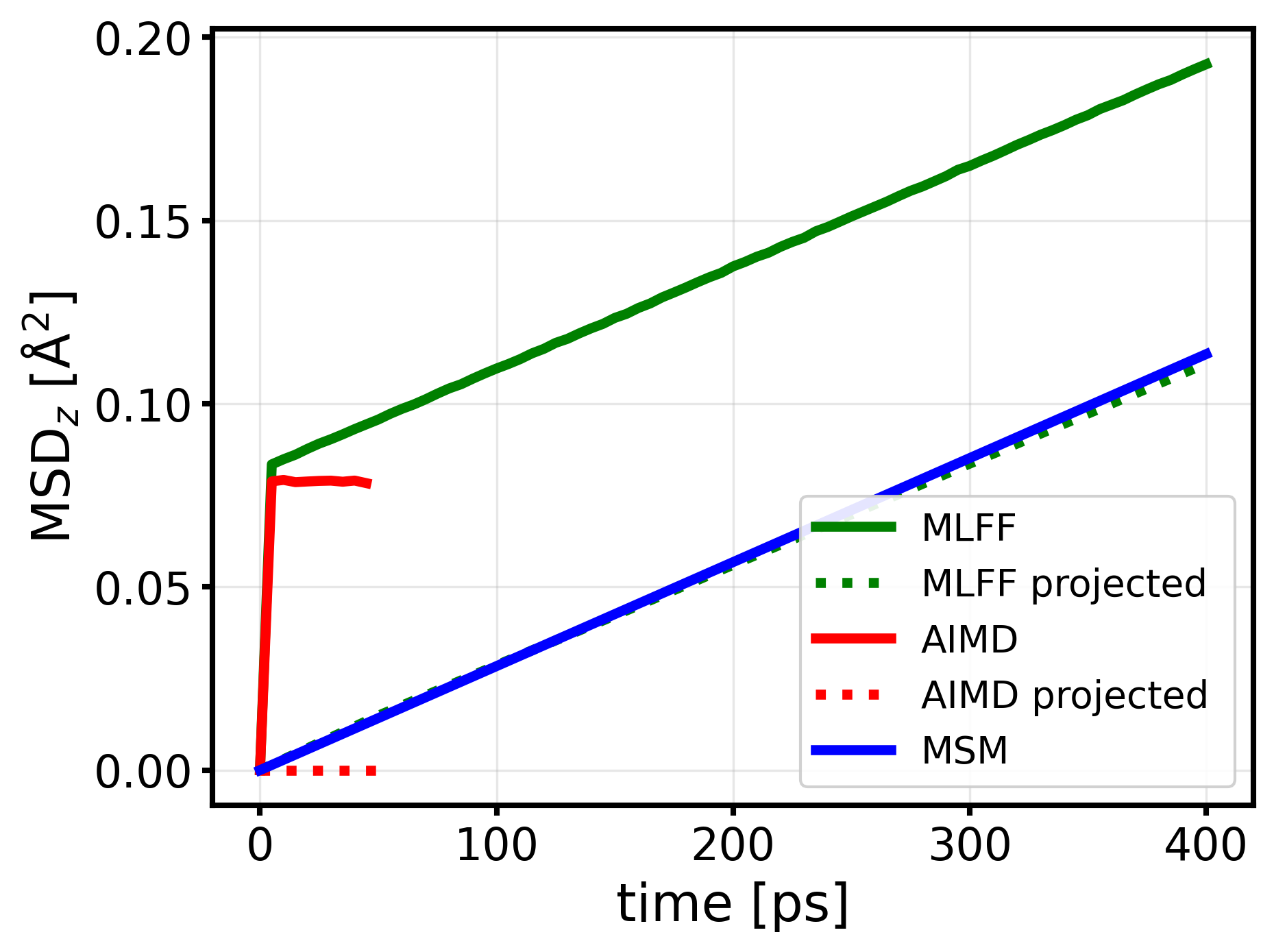}
\caption{$z$-component of the MSD of Li$_{13}$Si$_4$ obtained from 100~ps AIMD, 30~ns MLFF-MD, and Markov state model (MSM) simulations. For the ``projected'' MSDs, the positions of the Li atoms were mapped in each time step to the nearest lithium lattice sites. This procedure removes the contributions from local fluctuations around crystallographic lattice sites and is implicitly included in the Markov models by construction. Meaningful diffusion coefficients in z-direction can only be obtained from the MLFF and MSM simulations.}
\label{fig:z_direc}
\end{figure}

However, the situation becomes more complex when examining the directional components of the MSD. In Li$_{13}$Si$_4$, lithium mobility is dominated by diffusion along one-dimensional channels oriented along the $x$-direction, whereas diffusion along the $y$- and $z$-directions constitutes a rare event on AIMD timescales. Figure~\ref{fig:z_direc} compares the MSD along the $z$-direction obtained from short AIMD and long MLFF simulations. Meaningful slopes of the MSD, and therefore reliable diffusion coefficients, are accessible only from the MLFF simulations. The diffusion coefficient along the $x$-direction ($0.04\,\text{\AA}^2\text{ps}^{-1}$) is approximately 800 times larger than that along the $z$-direction ($5\times 10^{-5}\,\text{\AA}^2\text{ps}^{-1}$). Diffusion in $y$- and $z$-directions is not an artifact of the MLFF simulations as lithium diffusion in these directions can also be detected in AIMD simulations, as reported in Ref.~\citenum{kirsch2022atomistic}, but only at significantly elevated temperatures that help to overcome the higher migration barriers.

Motivated by the quasi-one-dimensional diffusion channel in Li$_{13}$Si$_4$ at 500\,K (cf.~Fig.~\ref{fig:jumps_six}),~\cite{kirsch2022atomistic} we performed a systematic size study that is impractical with AIMD but feasible with MLFFs. Elongating the supercell along the transport axis reveals that small boxes (e.g., $1\times 1\times 1$) overestimate the MSD slope due to unphysical correlation between the motion of an ion and the corresponding creation of a vacancy in the adjacent periodic image.
As the box is extended (e.g., $2\times 1\times 1$, $3\times 1\times 1$, $8\times 1\times 1$), the apparent diffusion coefficient $D$ decreases and the MSD curves converge to an asymptotic limit, indicating size convergence (Fig.~\ref{fig:msd_aimd_mlff}b). Consequently, the converged value of the diffusion coefficient obtained from the extended $8\times 1\times 1$ system ($0.0029\,\text{\AA}^2\text{ps}^{-1}$ at 500\,K) is in very good agreement with the experimental value $(0.0028-0.0051\,\text{\AA}^2\text{ps}^{-1}$ at 688\,K). In contrast, the diffusion coefficient calculated for the $1\times 1\times 1$ system is much higher ($0.042\,\text{\AA}^2\text{ps}^{-1}$ at 500\,K).\cite{Wen1981}

Hence, (i) AIMD-level diffusivities in anisotropic networks are biased high by finite-size effects, and (ii) MLFFs enable the larger simulation cells required to obtain size-independent transport coefficients. These converged trajectories provide the well-sampled jump statistics used to construct the Markov models discussed below.

\subsubsection{Lithium Jump Statistics and Markov State Models}

Transition matrices $\mathcal{M}^\tau$ were sampled from the MLFF trajectories according to eq.~\ref{eq:count}. By using the crystallographic Li positions as lattice sites, continuous MLFF trajectories are mapped onto discrete networks. A detailed protocol for the construction of the transition matrices was given in section~\ref{sec:create_markov}.

\begin{figure}[htbp]
  \centering
  \begin{subfigure}[t]{0.32\textwidth}
    \centering
    \includegraphics[width=\textwidth]{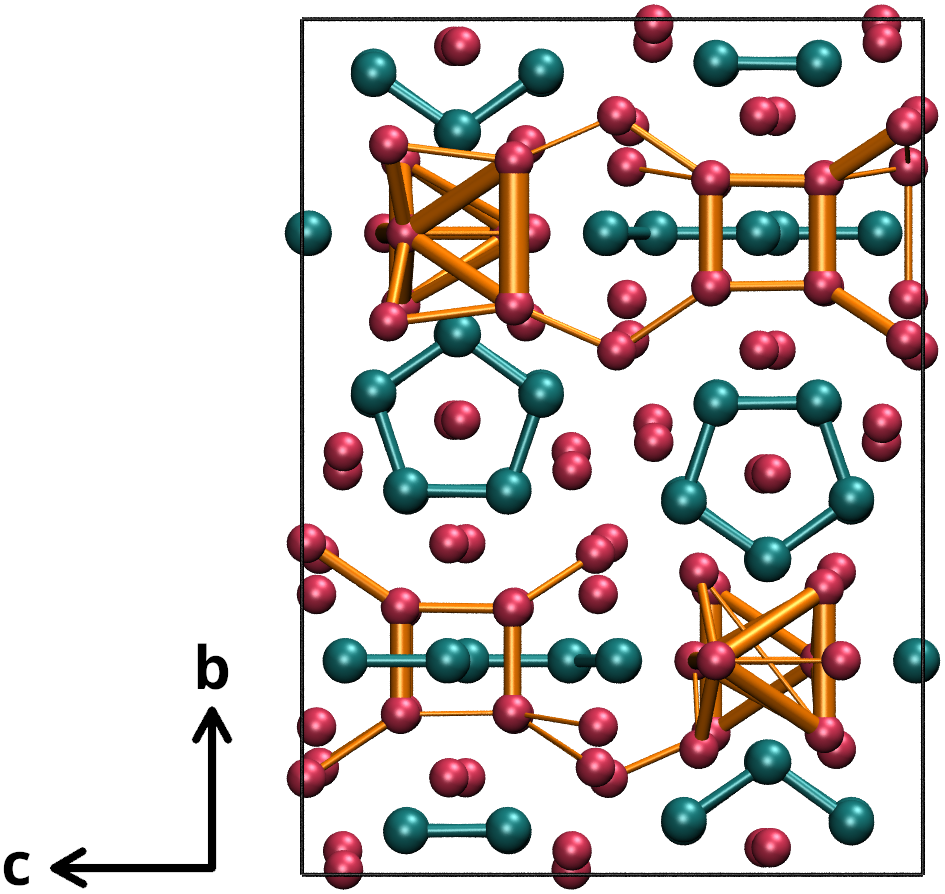}
    \caption{Li$_{12}$Si$_7$: AIMD (100 ps)}
  \end{subfigure}\hfill
  \begin{subfigure}[t]{0.32\textwidth}
    \centering
    \includegraphics[width=\textwidth]{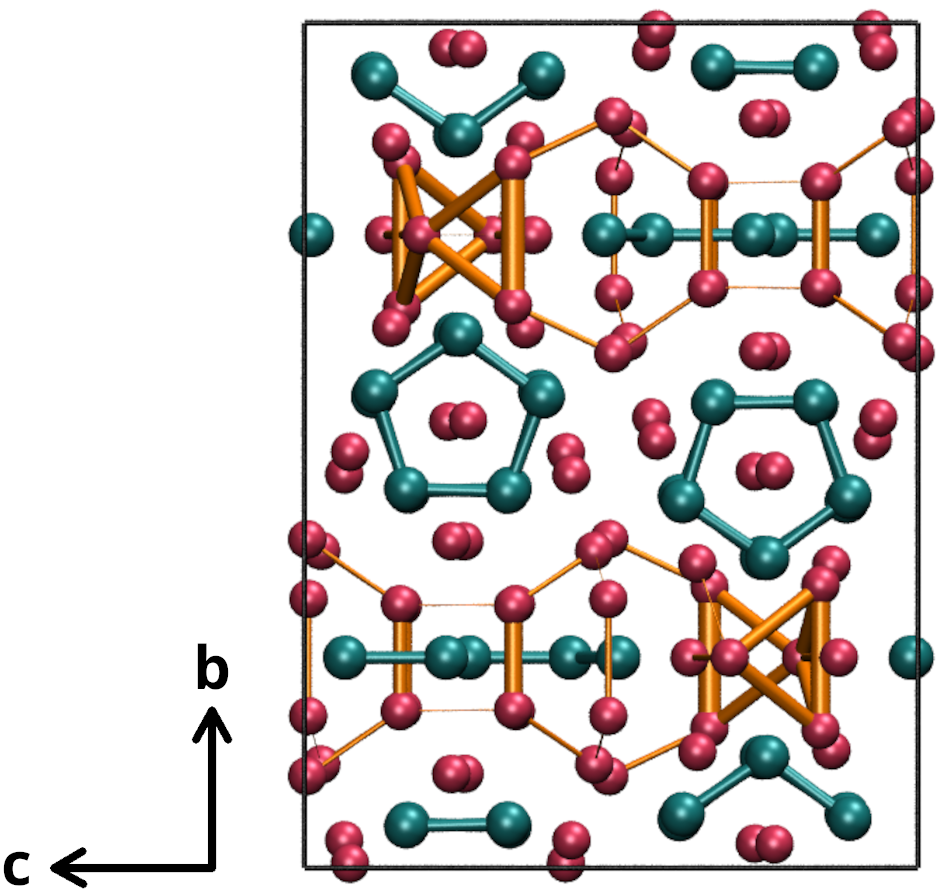}
    \caption{Li$_{12}$Si$_7$: fine-tuned MLFF (1 ns)}
  \end{subfigure}\hfill
  \begin{subfigure}[t]{0.32\textwidth}
    \centering
    \includegraphics[width=\textwidth]{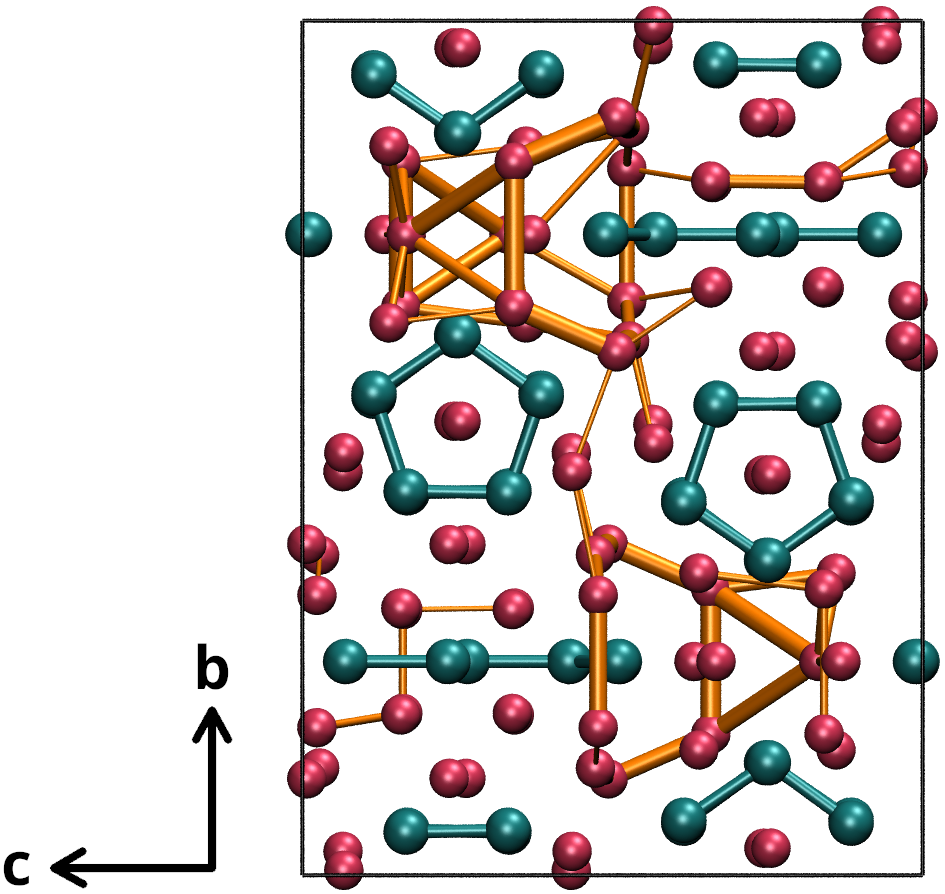}
    \caption{Li$_{12}$Si$_7$: foundation model (1 ns)}
  \end{subfigure}

  \vspace{0.6em}
  \begin{subfigure}[t]{0.32\textwidth}
    \centering
    \includegraphics[width=\textwidth]{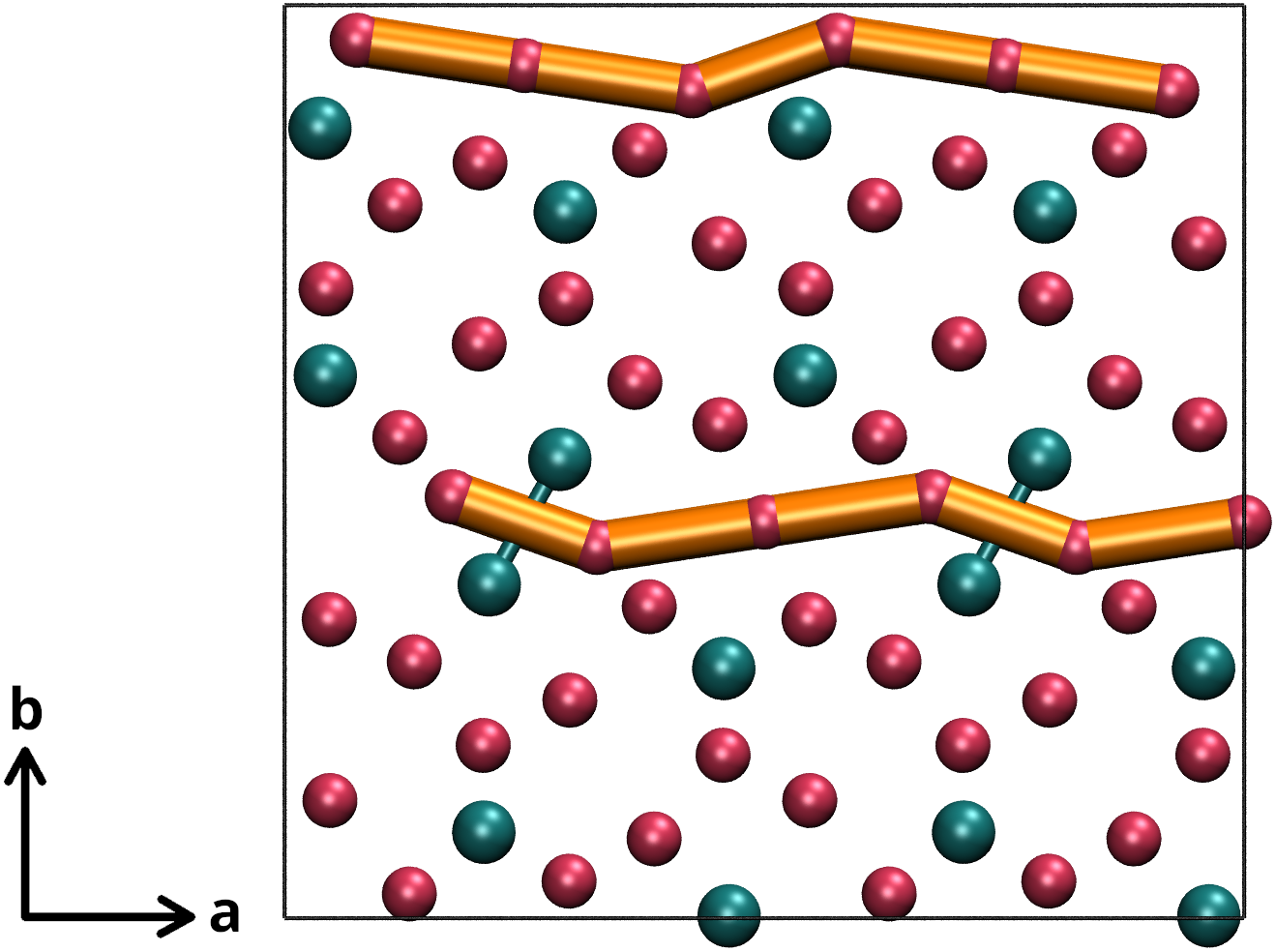}
    \caption{Li$_{13}$Si$_4$: AIMD (100 ps)}
  \end{subfigure}\hfill
  \begin{subfigure}[t]{0.32\textwidth}
    \centering
    \includegraphics[width=\textwidth]{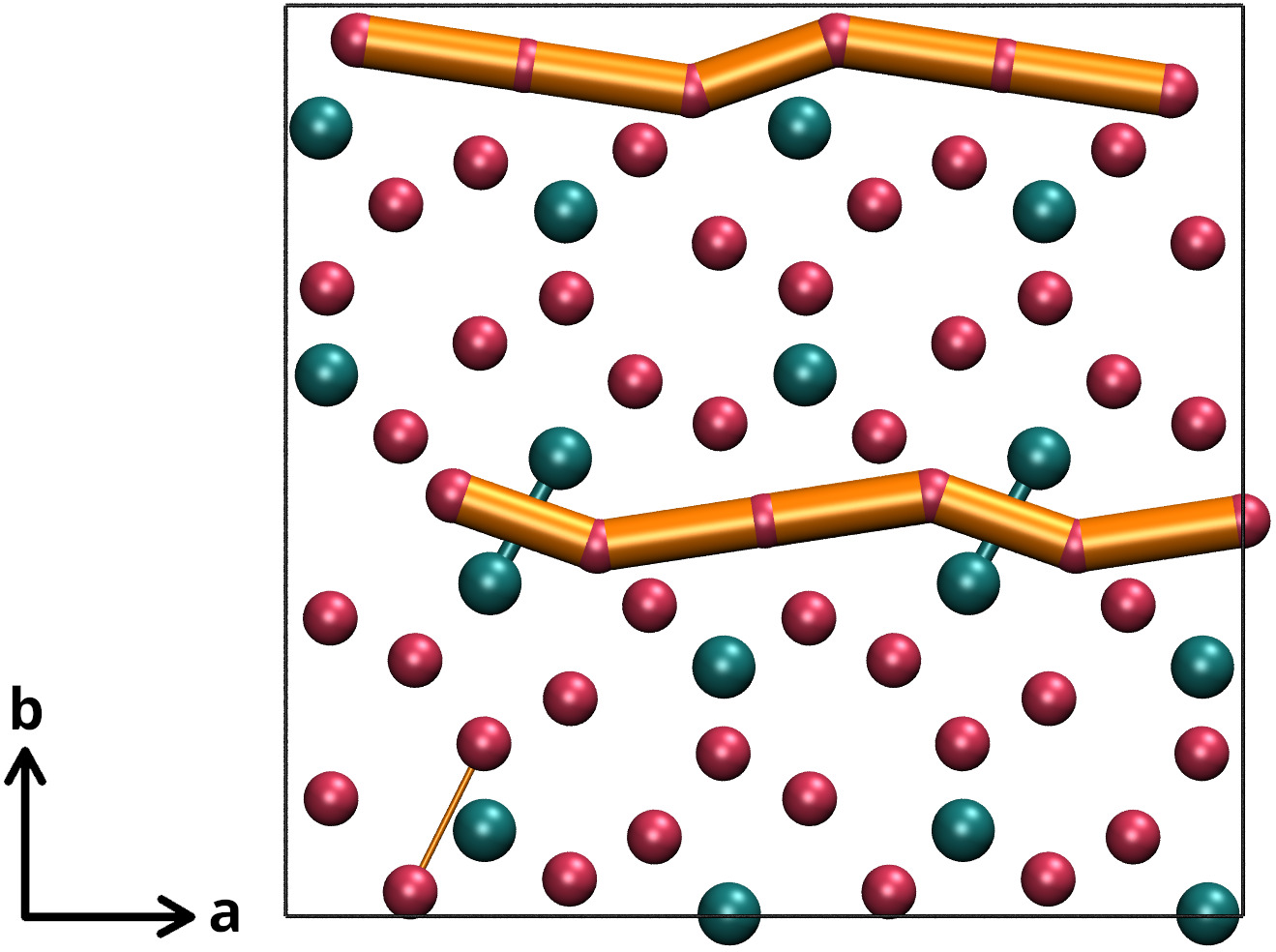}
    \caption{Li$_{13}$Si$_4$: fine-tuned MLFF (1 ns)}
  \end{subfigure}\hfill
  \begin{subfigure}[t]{0.32\textwidth}
    \centering
    \includegraphics[width=\textwidth]{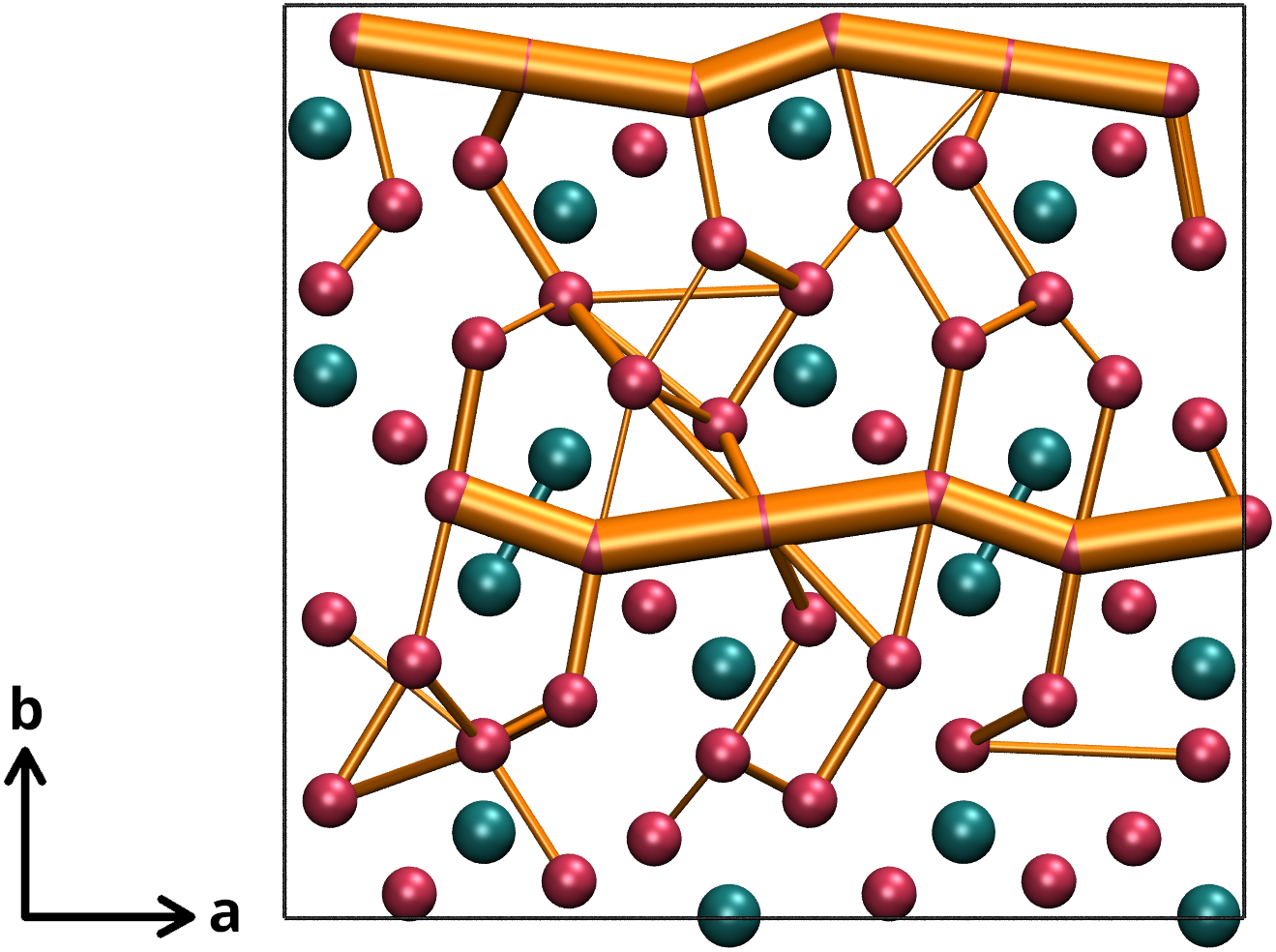}
    \caption{Li$_{13}$Si$_4$: foundation model (1 ns)}
  \end{subfigure}
  \caption{Average Li$^{+}$ jump frequency maps at 500~K. 
  Top row: Li$_{12}$Si$_7$ with (a) AIMD (100~ps), (b) fine-tuned MLFF (1~ns), (c) foundation model (1~ns). 
  Bottom row: Li$_{13}$Si$_4$ with (d) AIMD (100~ps), (e) fine-tuned MLFF (1~ns), (f) foundation model (1~ns). 
  The foundation model overestimates jump rates; fine-tuned MLFFs recover AIMD-observed channels and sample rare events.}
\label{fig:jumps_six}
\end{figure}

Figure~\ref{fig:jumps_six} shows representative jump networks, i.e., visualizations of the transition matrices $\mathcal{M}^{50\,\mathrm{fs}}$ for Li$_{12}$Si$_7$ and Li$_{13}$Si$_4$, extracted from AIMD (100~ps), fine-tuned MLFF (1~ns), and foundation-model (1~ns) trajectories. Line thickness encodes hop frequency. AIMD and fine-tuned MLFFs are in good agreement, which is expected as energy barriers for lithium jumps and diffusion coefficients were already in good agreement. The dominant one-dimensional diffusion in Li$_{13}$Si$_4$ becomes immediately apparent from Fig.~\ref{fig:jumps_six}d and Fig.~\ref{fig:jumps_six}e. The jump networks resulting from the foundation model are shown in Fig.~\ref{fig:jumps_six}c and Fig.~\ref{fig:jumps_six}f, and they are artificially dense due to underestimated migration barriers.

\begin{figure}[htbp]
  \centering
    \begin{minipage}{0.66\textwidth}
      \begin{minipage}{\textwidth}
       { \centering
      \includegraphics[width=0.9\textwidth]{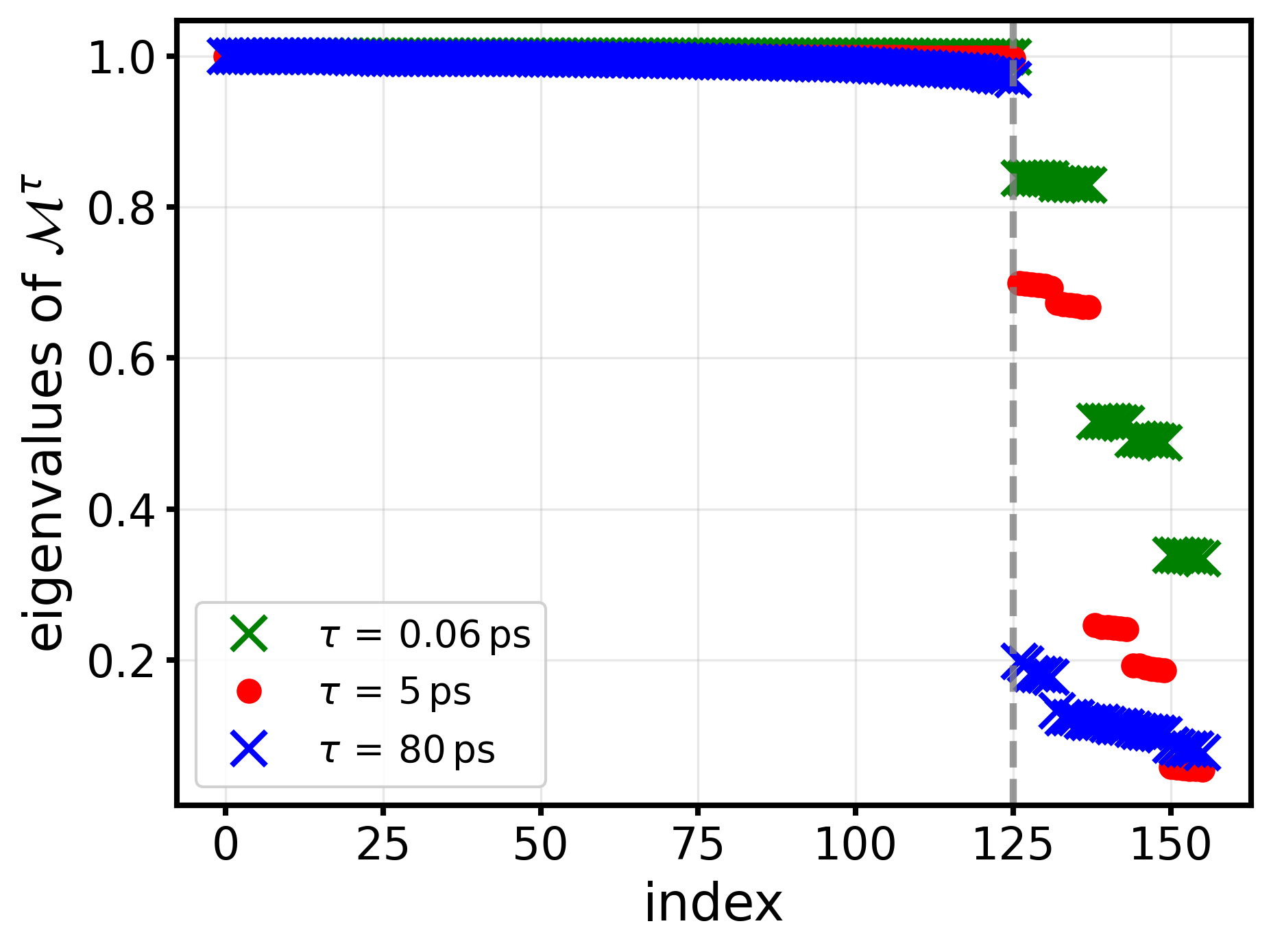}\\[-0.2em]}
       (a)
      \end{minipage}
      \begin{minipage}{\textwidth}
       { \centering
        \includegraphics[width=0.99\textwidth]{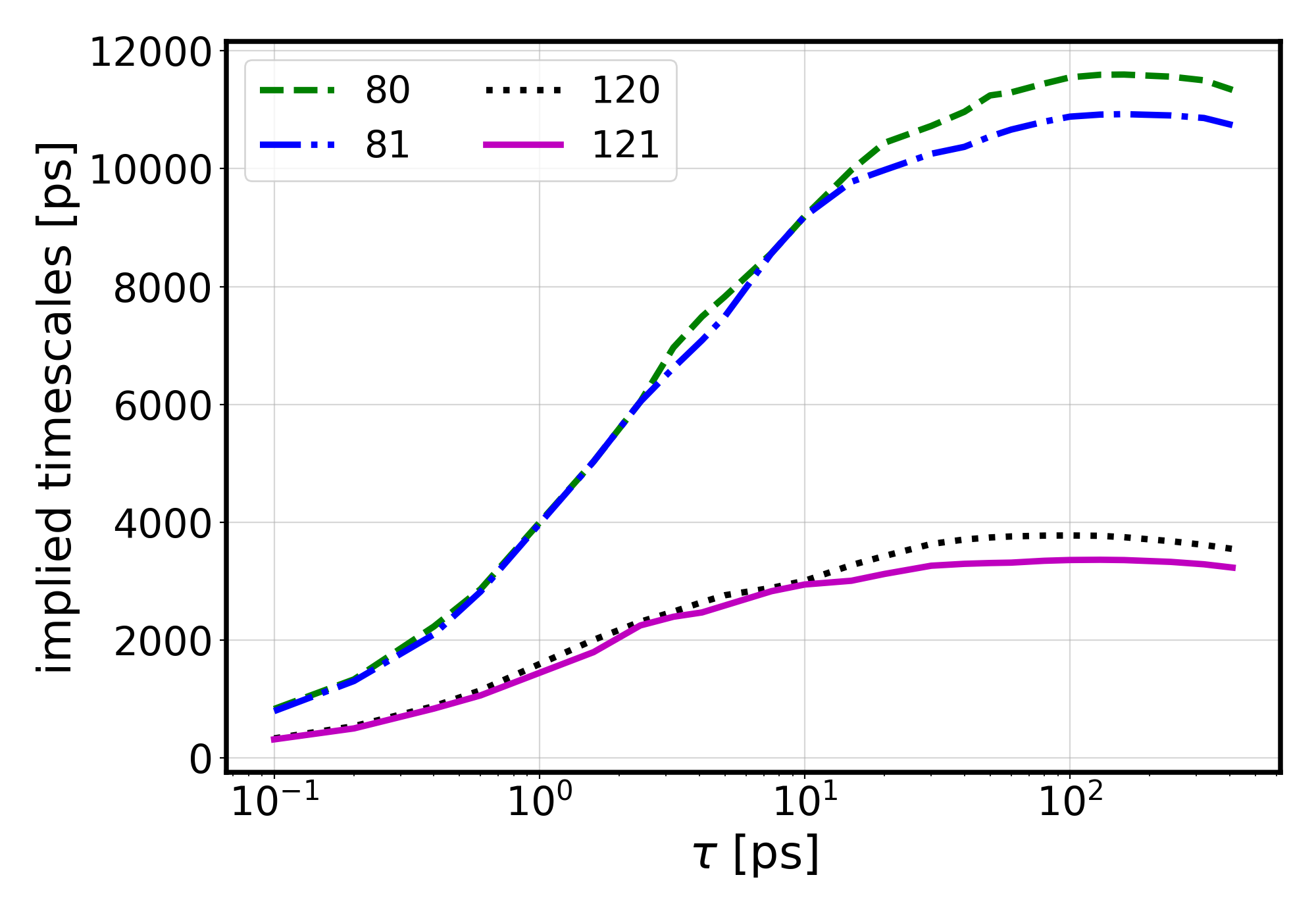}\\}
      (b) 
      \end{minipage}
    \end{minipage}
  \begin{minipage}{0.32\textwidth}    
    \begin{minipage}{0.9\textwidth}
      \centering
      \includegraphics[width=\textwidth]{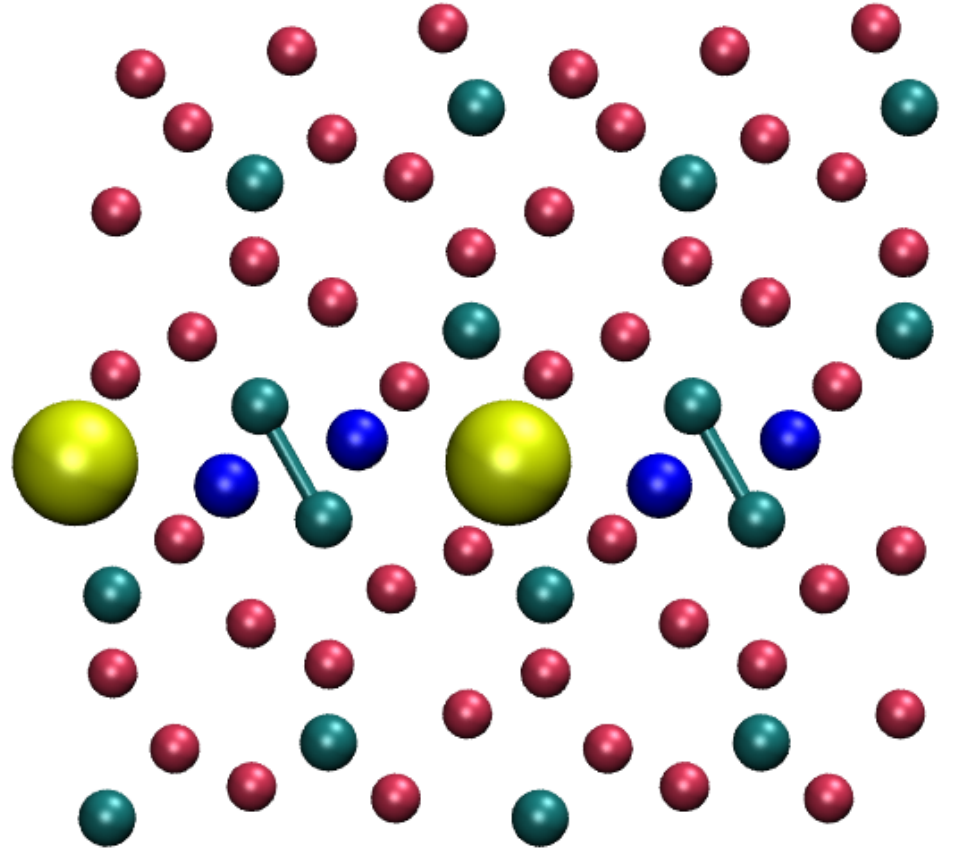}\\[-0.2em]
      \small (c) 155th eigenvector 
    \end{minipage}
    \begin{minipage}{0.9\textwidth}
      \centering
     
    \includegraphics[width=\textwidth]{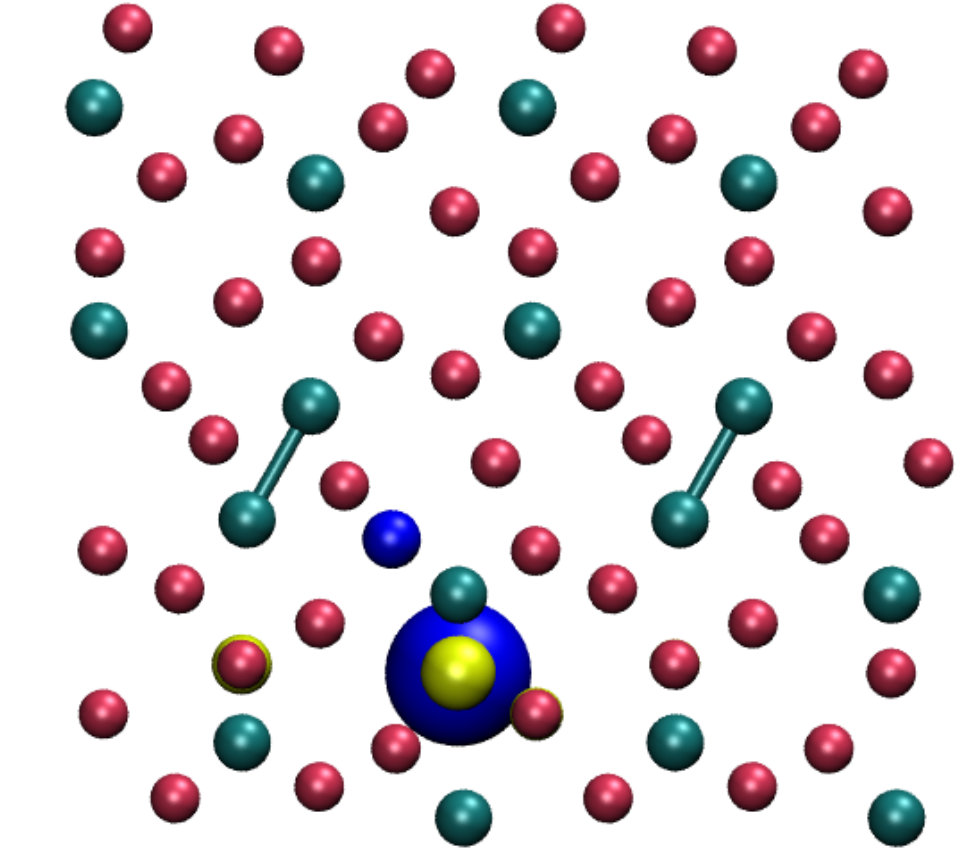}\\[-0.2em]
      \small (d) 122nd eigenvector 
    \end{minipage}
    \begin{minipage}{0.9\textwidth}
      \centering
      \includegraphics[width=\textwidth]{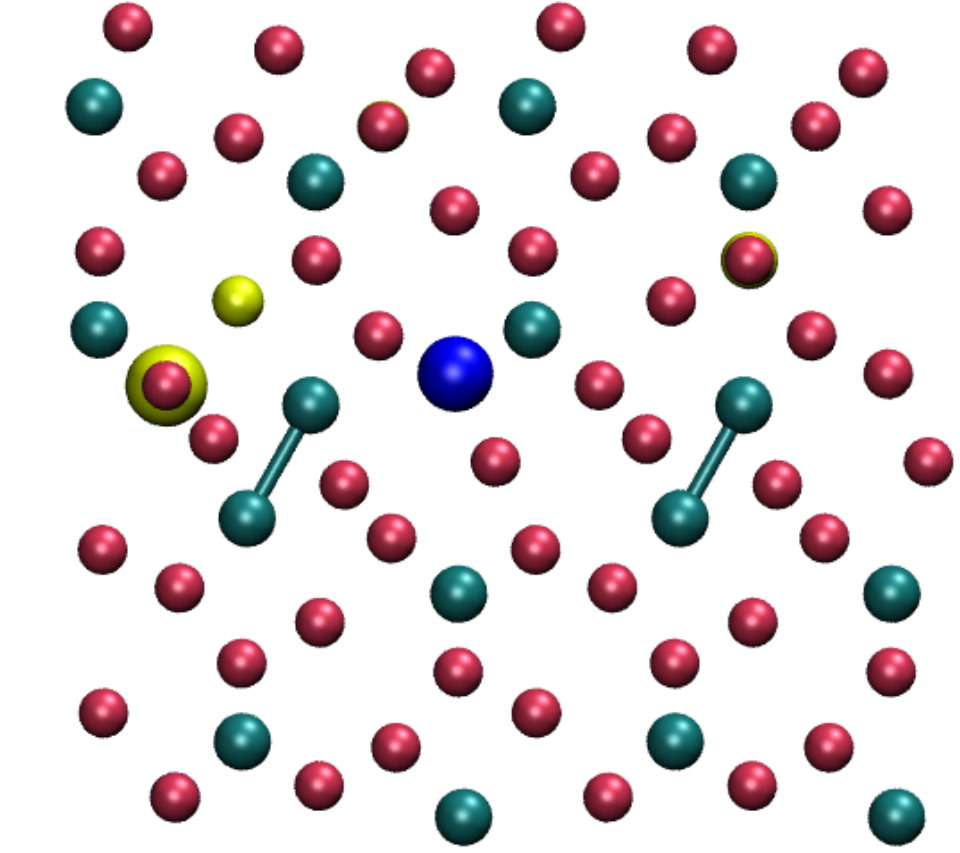}\\[-0.2em]
      \small (e) 81st eigenvector
    \end{minipage}
  \end{minipage}
 \caption{(a) Eigenvalue spectrum of the transition matrix $\mathcal{M}^{\tau}$ for Li$_{13}$Si$_4$ at $\tau \in \{0.06,\, 5,\, 80\,\mathrm{ps}\}$. 
(b) Implied timescales $t_k(\tau) = -\tau / \ln \lambda_k(\tau)$ calculated from the eigenvalues of $\mathcal{M}^\tau$ for different lag times $\tau$. 
(c)--(e) Visualization of selected eigenvectors.}
  \label{fig:markov_implied}
\end{figure}

In addition, we systematically analyzed the transition matrix $\mathcal{M}^{\tau}$ for different lag times $\tau$ in terms of its eigenvalues and eigenvectors. The absolute value of each eigenvalue lies between 0 and 1. The eigenvalue spectrum of $\mathcal{M}^{\tau}$ for Li$_{13}$Si$_4$ at $\tau \in \{0.06,\, 5,\, 80\,\mathrm{ps}\}$ is shown in Fig.~\ref{fig:markov_implied}a. Since the simulation cell contained 156 Li atoms, 156 eigenvalues were obtained. The eigenvalues can be grouped into two distinct regions: the first 124 are very close to unity, while the remaining 32 are significantly smaller. The largest eigenvalue equals one, and its corresponding eigenvector describes the uniform distribution of lithium atoms across the available lattice sites. This state represents the equilibrium distribution expected from the MD trajectory. Eigenvectors associated with smaller eigenvalues describe relaxation processes toward this equilibrium distribution. 

Three representative eigenvectors of the transition matrix $\mathcal{M}^{80\,\mathrm{ps}}$ for Li$_{13}$Si$_4$ are visualized in Figs.~\ref{fig:markov_implied}c--e. For our specific choice of crystallographic Li positions as lattice sites, these eigenvectors can be correlated with lithium diffusion pathways. Eigenvectors corresponding to eigenvalues significantly smaller than one (index $\geq 125$) exhibit non-zero components only for Li atoms located within a single one-dimensional diffusion channel. A particular example is the 155th eigenvector shown in Fig.~\ref{fig:markov_implied}c. These delocalized eigenvectors represent fast, collective ``rattling'' motions of Li atoms within the one-dimensional channels. In contrast, eigenvectors with indices smaller than 125 are more spatially localized and also contribute to displacements of Li atoms along the $y$- and $z$-directions. These eigenvectors are more complex and cannot be classified as simple channel modes. Representative examples are the 81st and 122nd eigenvectors (Figs.~\ref{fig:markov_implied}d and \ref{fig:markov_implied}e).
Diffusion along the $y$- and $z$-directions is not an artifact. Li jumps in these directions occur with much lower frequency compared to those along the $x$-direction. Consequently, the line thicknesses representing Li jump frequencies between lattice sites in Fig.~\ref{fig:jumps_six} are too small to be visible. A similar observation was made in Ref.~\citenum{kirsch2022atomistic}, where Li$_{13}$Si$_4$ at 800~K also exhibited a measurable number of Li jumps in the $y$- and $z$-directions. 

Figure~\ref{fig:markov_implied}b shows the implied timescales, defined in eq.~\ref{eq:tau_conv}, as a function of lag time $\tau$ for four representative eigenvalues. Different eigenvalues exhibit distinct minimal lag times $\tau_\mathrm{min}$, marking the onset of the plateau region in the implied-timescale curves. Smaller eigenvalues correspond to faster relaxation processes and therefore reach their plateau at shorter lag times. In the Supporting Information, we show the implied timescales for all eigenvalues. 
Implied timescales of eigenvalues very close to one do not fully reach the plateau region even after lag times greater than 100 ps.

\begin{figure}[htbp]
  \centering
  \begin{minipage}{0.49\textwidth}
    \centering
    \includegraphics[width=\textwidth]{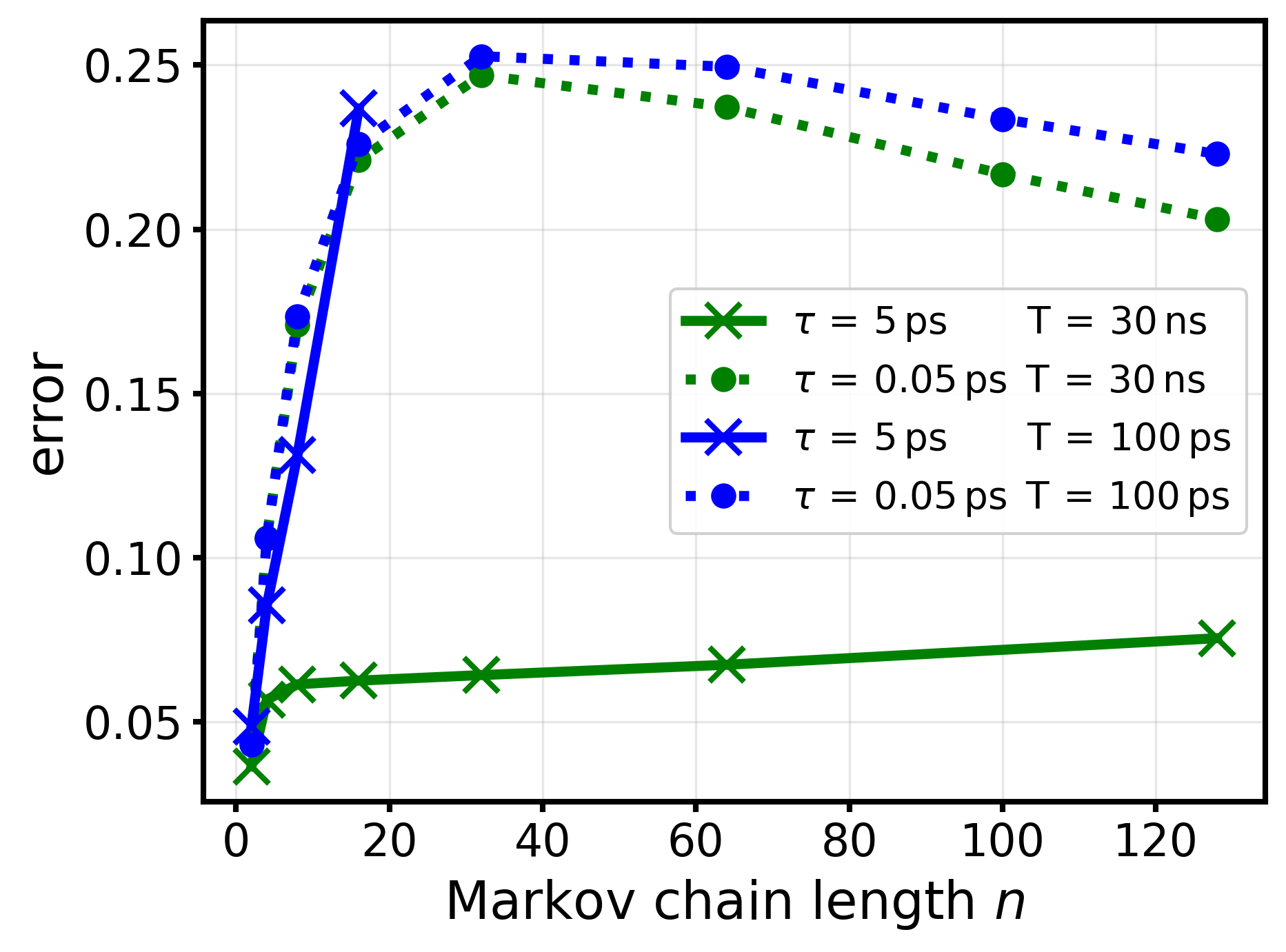}\\[-0.2em]
    \small (a) Li$_{13}$Si$_4$
  \end{minipage}
  \begin{minipage}{0.49\textwidth}
    \centering
    \includegraphics[width=\textwidth]{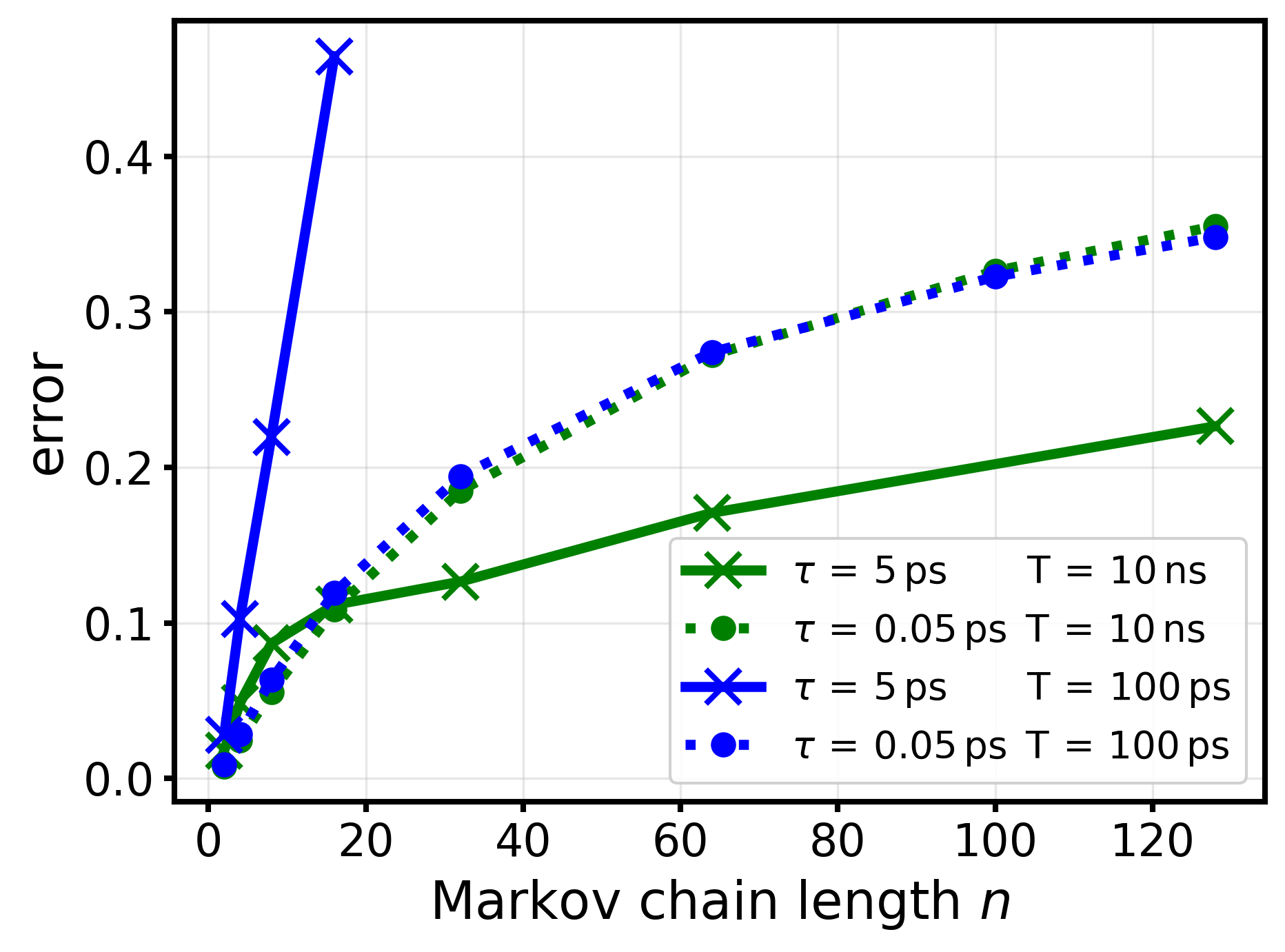}\\[-0.2em]
    \small (b)  Li$_{12}$Si$_7$
  \end{minipage}
  \caption{Evaluation of the Markov property using the Chapman--Kolmogorov test. The plot compares the directly sampled multi-lag transition matrix $\mathcal{M}^{n\tau}_{\mathrm{sampled}}$ with the Markovian prediction $(\mathcal{M}^\tau)^n$. The relative error $\mathrm{err}(n)$, defined in eq.~\ref{eq:err}, is shown on the $y$-axis as a function of the Markov chain length $n$. The solid blue line is shown only up to $n=16$ because transition matrices can be sampled from 100~ps trajectories only up to a maximum lag time of $\tau \cdot n = 5\,\mathrm{ps}\cdot 16 = 80\,\mathrm{ps}$.}
  \label{fig:markov_err}
\end{figure}

Finally, the Chapman--Kolmogorov test (eq.~\ref{eq:chap}) was used to determine the minimal lag time $\tau$ required for constructing a consistent Markov model. The relative error $\mathrm{err}(n)$, defined in eq.~\ref{eq:err}, quantifies the deviation between the directly sampled multi-lag transition matrix $\mathcal{M}^{n\tau}_{\mathrm{sampled}}$ and the Markovian prediction $(\mathcal{M}^\tau)^n$ (Fig.~\ref{fig:markov_err}). We systematically analyzed $\mathrm{err}(n)$ as a function of (i) the lag time $\tau$ and (ii) the total trajectory length $T$. Lag times of $\tau = 5$~ps (comparable to $\tau_{\mathrm{min}}$) and $\tau = 0.05$~ps (well below $\tau_{\mathrm{min}}$) were examined, together with trajectory lengths of 100~ps (typical of AIMD) and 10--30~ns (typical of MLFF-MD).  

Note that error plots for different $\tau$ share the same $x$-axis (the Markov chain length $n$) but correspond to different physical times $n\tau$. For example, a chain length of $n=100$ represents a physical time of 5~ps for $\tau=0.05$~ps and 500~ps for $\tau=5$~ps.  

For long MLFF trajectories ($T \approx 10$~ns) and lag times near $\tau_{\mathrm{min}}$, the CK error remains very small. In these cases, transition matrices sampled at $\tau = 5$~ps accurately reproduce Li$^+$ dynamics over time intervals up to $128\times\tau$ ($\sim$600~ps) with an error of only $\sim$15\%. In contrast, MSMs constructed from short (100~ps) trajectories exhibit substantially larger errors, highlighting the need for extended sampling to achieve Markovian consistency. 

\begin{figure}[htbp]
  \centering
  \includegraphics[width=0.8\linewidth]{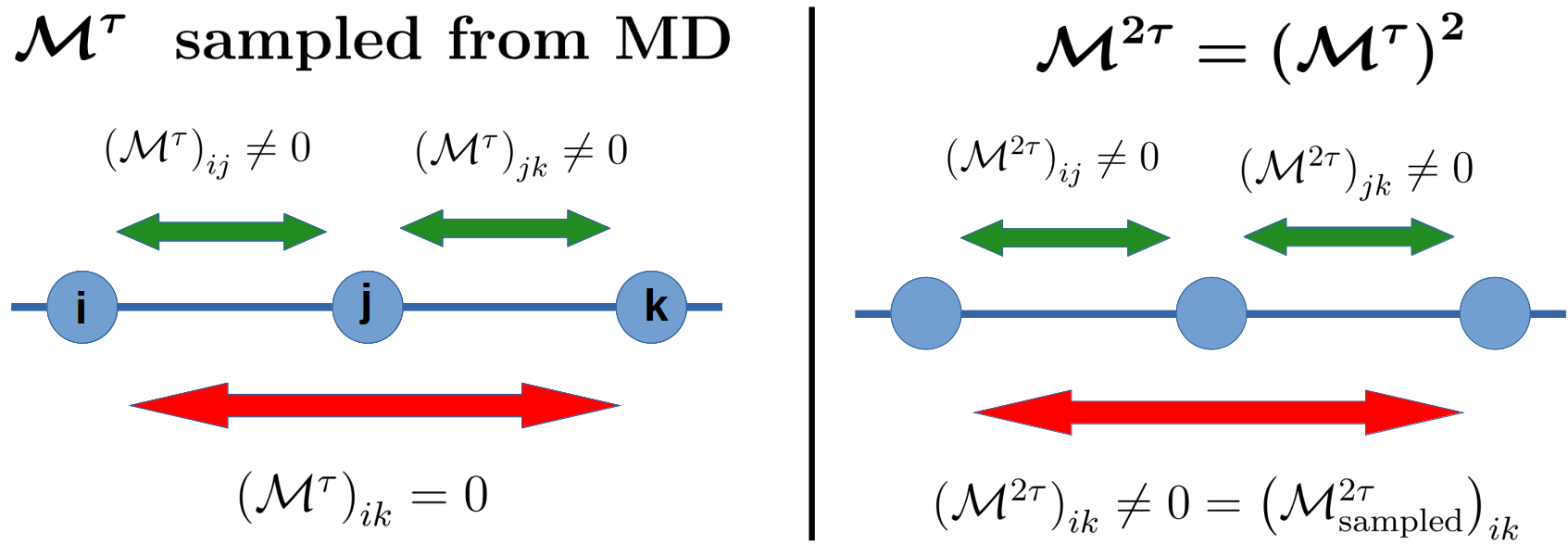}
  \caption{Illustration of how the discretization of the state space imposes a lower bound on the lag time $\tau$ required for constructing consistent MSMs. Sampling transition matrices $\mathcal{M}^\tau$ at very short lag times results in non-zero matrix elements describing Li transfer probabilities between neighboring lattice sites, while matrix elements corresponding to next-nearest neighbors remain zero. The transition matrix $\mathcal{M}^{2\tau}$ for the interval $2\tau$ is obtained by squaring $\mathcal{M}^\tau$. By construction, this matrix now includes non-zero elements that represent ion transfer probabilities between next-nearest neighbors, even if no such jumps were directly detected in the MD trajectory within the interval $2\tau$.}
  \label{fig:mini_tau}
\end{figure}

For lag times shorter than $\tau_{\mathrm{min}}$, we observe comparable errors for transition matrices sampled from both short (100~ps) and long (10--30~ns) trajectories. This indicates that at very short lag times, AIMD-level sampling is already sufficient to capture local transitions. However, the errors for such short lag times (independent of the overall length of the trajectory $T$) display a pronounced non-monotonic dependence on the chain length $n$, reflecting the breakdown of the Markov assumption and the presence of unresolved memory effects in the underlying dynamics. Consequently, the observed non-Markovian behavior does not originate from insufficient sampling but rather from violating the minimum lag time required for proper state decorrelation, which is imposed by the discretization of the phase space. Figure~\ref{fig:mini_tau} illustrates how the discretization of the state space imposes a lower bound on the lag time $\tau$ required for constructing consistent MSMs.

\begin{figure}[htbp]
  \centering
  \begin{minipage}{0.49\textwidth}
    \centering
    \includegraphics[width=\textwidth]{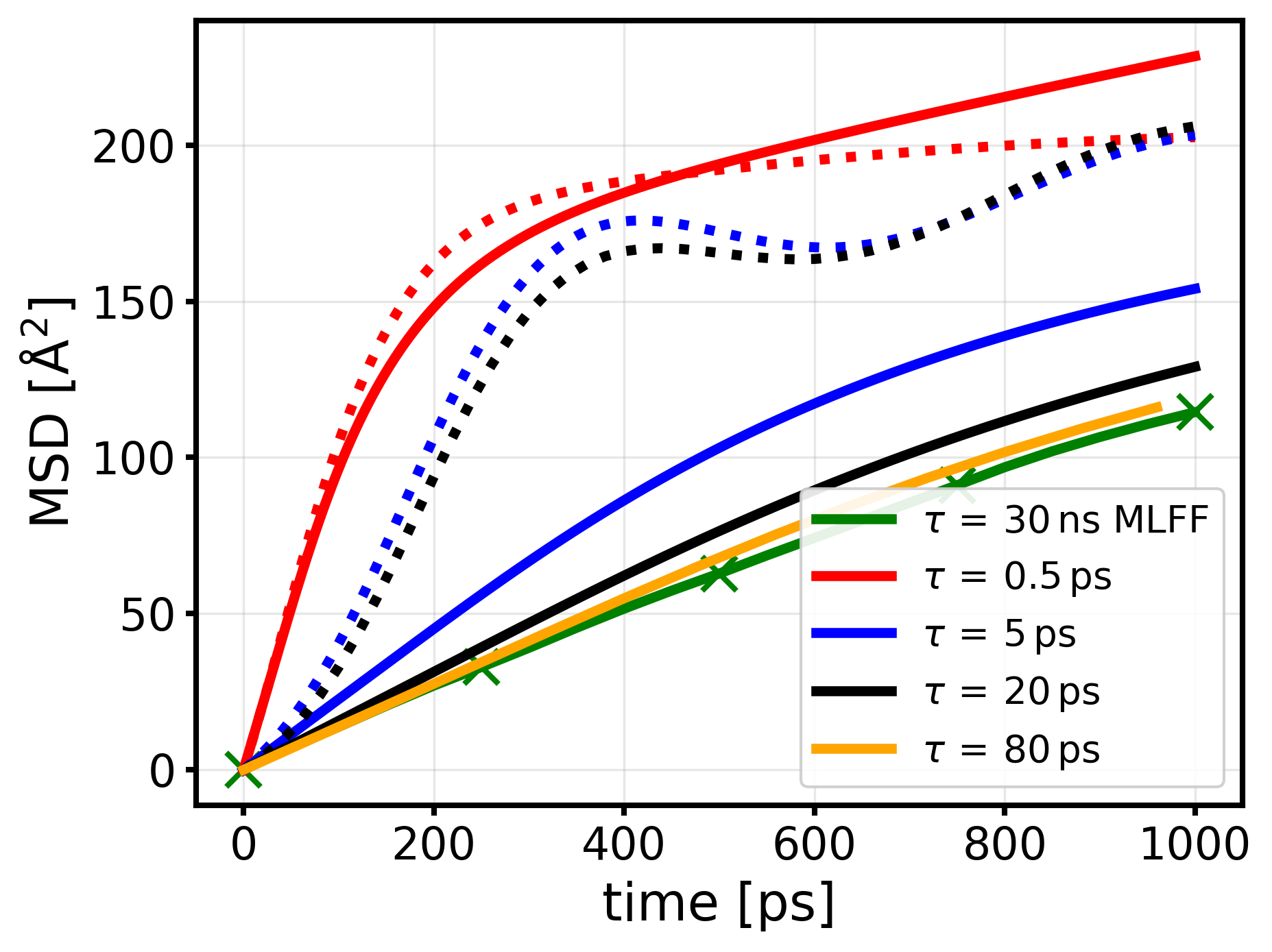}\\[-0.2em]
    \small (a) Li$_{13}$Si$_4$
  \end{minipage}
  \begin{minipage}{0.49\textwidth}
    \centering
    \includegraphics[width=\textwidth]{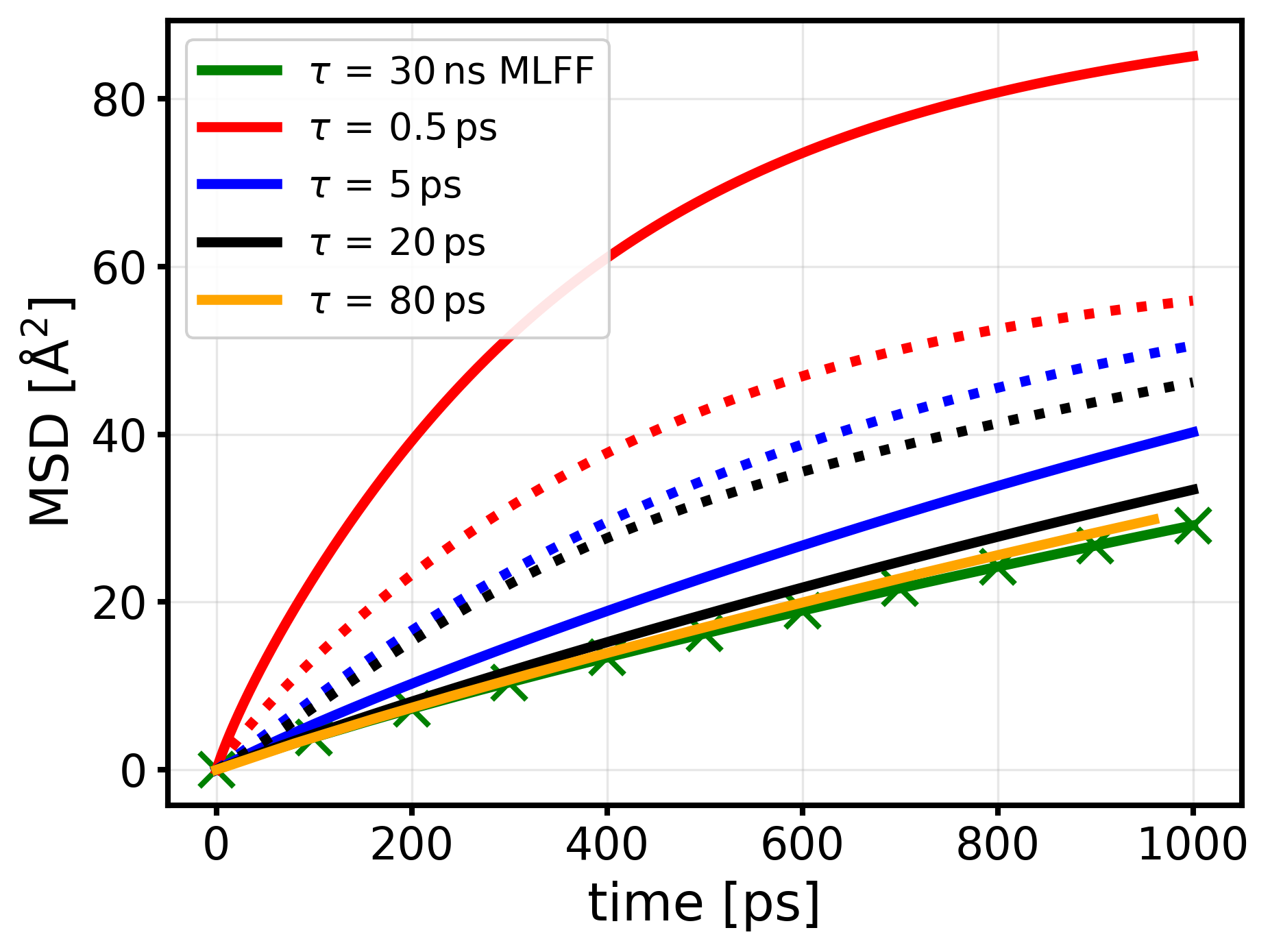}\\[-0.2em]
    \small (b) Li$_{12}$Si$_7$
  \end{minipage}
    \caption{MSD at 500~K from MLFF-MD (green) and MSM reconstructions from transition matrices $\mathcal{M}^{n\tau}$ sampled with lag time $\tau$. Solid (dashed) lines are obtained from transition matrices sampled from MLFF simulations of total length 30~ns (100~ps). MSMs trained on long trajectories with minimal lag times $\tau > 5~\textrm{ps}$ reproduce long-time behavior, while MSMs from 100~ps MD and shorter lag times deviate. Calculation of MSDs from transition matrices according to eq.~\ref{eq:msd} does not take into account periodic images. Therefore, the maximum displacement is half of the box dimensions, leading to a plateau on very long timescales for some of the models.}
  \label{fig:msd_markov}
\end{figure}

In the final step, transport properties were reconstructed by propagating displacements through the MSM according to eq.~\ref{eq:prop}. The mean-square displacement of the Li ions was calculated from the Markov chain using eq.~\ref{eq:msd}. Figure~\ref{fig:msd_markov} compares MSDs for Li$_{12}$Si$_7$ and Li$_{13}$Si$_4$ obtained from MSMs sampled at different lag times and trajectory lengths. MSDs for both compounds were calculated from transition matrices constructed for lag times of $\tau \in \{0.5,\, 5,\, 20,\, 80~\mathrm{ps}\}$. Sampling intervals of length $\tau$ were obtained from 100~ps trajectories (typical of AIMD) and from 10--30~ns trajectories (typical of MLFF-MD). For $\tau = 80~\mathrm{ps}$, transition matrices could only be constructed from the MLFF-MD trajectories.  

All MSMs trained on 100~ps trajectories overestimate long-time transport. Similarly, MSMs constructed with lag times shorter than 5~ps deviate significantly from direct MLFF-MD results, indicating a breakdown of the Markovian approximation. Only MSMs sampled from 10--30~ns MLFF trajectories and lag times greater than 5~ps accurately reproduce the MSDs up to 1~ns. The agreement between MSMs and direct MLFF results improves systematically with increasing lag time and is good, very good, and excellent for $\tau$ values of 5, 20, and 80~ps, respectively.  

As a representative example demonstrating the advantage of the extended timescales accessible through MLFF and MSM simulations compared to AIMD, Figure~\ref{fig:z_direc} shows the MSD along the $z$-direction for Li$_{13}$Si$_4$ obtained from 100~ps AIMD, 30~ns MLFF, and MSM simulations. The MSM was constructed from the transition matrix $\mathcal{M}^{80\,\mathrm{ps}}$ sampled from the long MLFF trajectory. For very short timescales ($<$10~ps), the MSD offset agrees well between the AIMD and MLFF simulations. However, the AIMD-derived MSD remains constant at longer times, whereas the slopes of the MSD curves from MLFF and MSM simulations show excellent agreement in this regime. The inability of AIMD to capture the correct long-time slope - and the strong consistency between the MLFF and MSM results - becomes even more evident when fluctuations around the crystallographic lattice sites are removed by projecting the Li positions at each timestep onto the nearest lattice site. The ``projected'' MSD obtained from AIMD is zero at all times, indicating the absence of Li jumps between lattice sites with different $z$-coordinates, whereas the projected MSDs from the MLFF and MSM simulations are nearly indistinguishable.

In summary, reliable MSMs for Li--Si systems require lag times greater than 5~ps and training trajectories on the order of tens of nanoseconds. Once constructed, these models accurately predict the collective dynamical evolution of all Li$^+$ ions through simple matrix--vector propagation. This approach extends accessible temporal and spatial scales and enables simulation of lithium mobility on the microsecond timescale. Such scales are sufficient to augment existing continuum and multiscale models of silicon anodes with a physically grounded description of lithium diffusion within these materials.~\cite{Zhang2017,Silveri2024,Xie2016,DiLeo2015,Zhao2011}  

Previous multiscale approaches have primarily focused on chemomechanical modeling of lithiation-induced failure in high-volume-change anodes, often employing diffusion coefficients that do not explicitly depend on the local lithium concentration. As a next step, we aim to couple MSMs sampled at different lithium concentrations - particularly from amorphous Li-Si systems - to derive concentration-dependent diffusion models applicable at the electrode scale.

 \section{Conclusions}

In this work, we present a robust multiscale framework that bridges quantum-accurate atomistic simulations with mesoscale lithium transport by integrating \textit{ab initio} molecular dynamics, machine-learned force fields, and Markov state models. This combination overcomes the temporal and spatial limitations of conventional atomistic simulations while retaining first-principles fidelity. Fine-tuned equivariant MLFFs reproduce DFT migration barriers within a few percent and enable nanosecond-scale simulations in large supercells, revealing anisotropic and collective lithium motion that remains inaccessible to AIMD. We further demonstrated that only such long-time simulations remove the strong finite-size bias inherent to AIMD and yield diffusion coefficients in very good agreement with experiment.

The long MLFF trajectories provide statistically converged lithium jump networks from which MSMs can be constructed. The discretization of the state space imposes a minimal lag time for consistent propagation; for Li$_{13}$Si$_4$ and Li$_{12}$Si$_7$, lag times of 5--20~ps are required to generate MSMs that accurately reproduce mean-square displacements and diffusivities, including rare diffusion events that never occur within AIMD-accessible timescales. The extended time and length scales afforded by MLFF molecular dynamics are therefore essential, as they provide sufficiently many statistically independent intervals at the required lag times.

The resulting MSMs remain Markovian over more than two orders of magnitude in the lag times at which they were sampled, as verified through Chapman-Kolmogorov tests. Beyond enabling efficient stochastic propagation, the transition matrices offer mechanistic insight: eigenvalues and eigenvectors of the MSM encode characteristic relaxation timescales and spatially resolved diffusion modes. We showed that these eigenvectors correlate directly with lithium diffusion pathways, allowing a compact spectral representation of transport processes.

Although demonstrated here for defect-free crystalline Li-Si phases, the AIMD$\rightarrow$MLFF$\rightarrow$MSM workflow is general and extensible, especially in view of more realistic systems, such as amorphous systems. The local lithium coordination environments and migration mechanisms identified in crystalline phases constitute fundamental building blocks for diffusion in disordered systems. In amorphous Li-Si, the same elementary lithium jumps occur within a structurally heterogeneous network that introduces distributions of barrier heights and correlated motion. Extending our approach by training MLFFs on amorphous configurations and defining MSM states through geometric clustering will enable quantitative predictions of how structural disorder influences macroscopic diffusivity - an essential requirement for optimizing amorphous silicon anodes in practical batteries. 

In future work, we aim to apply this framework across a broader range of lithium silicide compositions (Li$_{15}$Si$_4$, Li$_{17}$Si$_4$, LiSi, etc.), temperatures, and defect configurations to systematically determine how structure, composition, and lithium concentration govern transport throughout the Li-Si system.

\subsection{Data Availability}

The MACE models, simulation scripts, and datasets used in this study will be made publicly available upon publication via an open-access repository. In the meantime, they are available from the corresponding author upon reasonable request.

\subsection{Acknowledgements}
The authors thank the staff of the Compute Center of the Technische Universität Ilmenau and especially Mr. Henning Schwanbeck for providing an excellent research environment. This work is supported by the Deutsche Forschungsgemeinschaft DFG (Project 435886714).




\FloatBarrier

\clearpage

\bibliography{paper}

\clearpage
\appendix

\includepdf[pages=-]{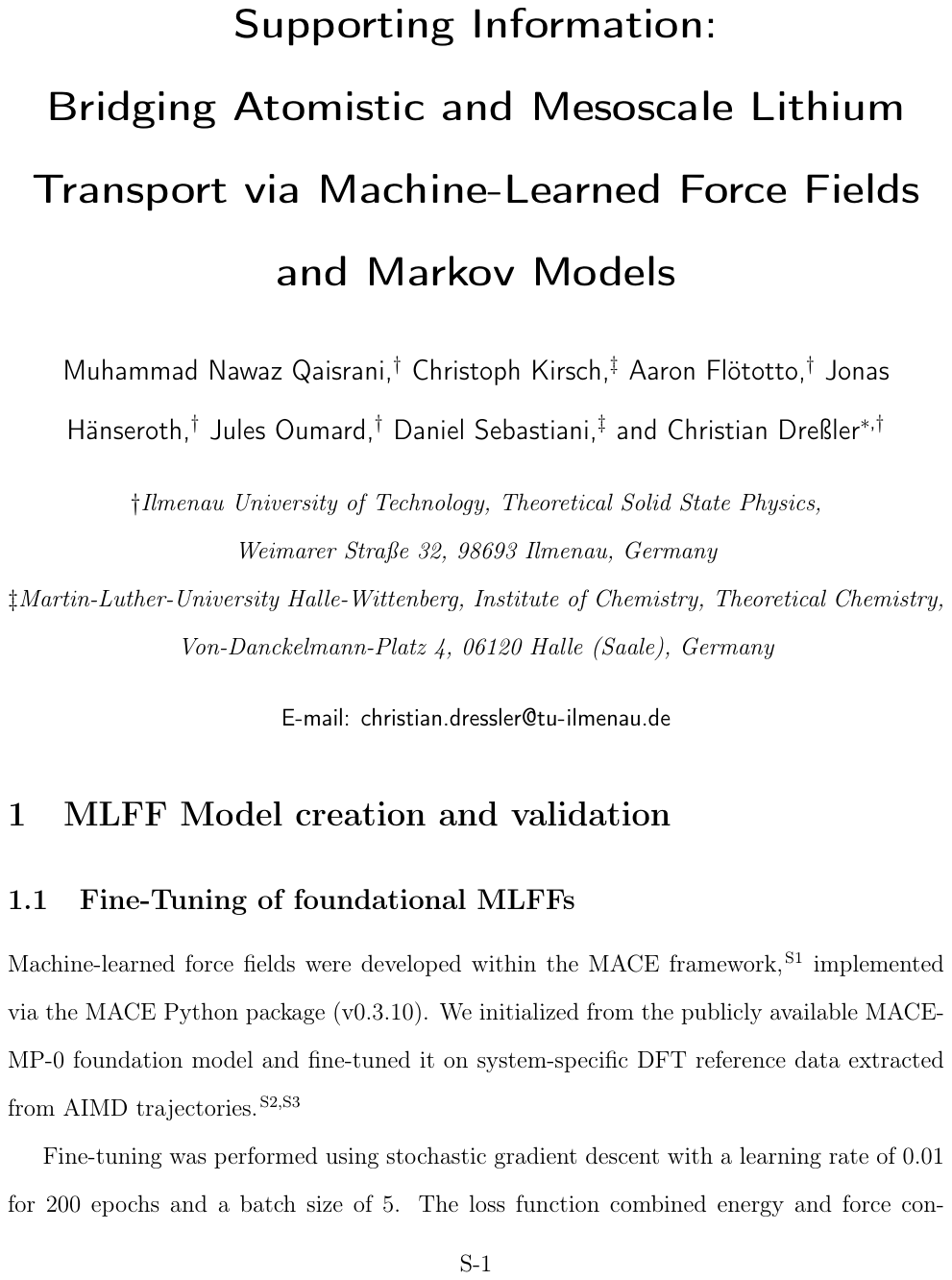}

\end{document}